\pdfoutput=1

\documentclass[11pt]{article}

\usepackage[]{acl}

\usepackage{times}
\usepackage{latexsym}
\usepackage{url}
\usepackage{subfigure}
\usepackage{xcolor}

\usepackage[T1]{fontenc}

\usepackage[utf8]{inputenc}

\usepackage{microtype}

\usepackage{inconsolata}
\usepackage{makecell}
\usepackage{float}
\usepackage{bm}
\usepackage{booktabs}
\usepackage{multirow}
\usepackage{colortbl}
\usepackage{xspace}
\usepackage{graphicx} 
\usepackage{color}
\usepackage{paralist}
\usepackage{amsmath,amsfonts,amsthm}

\title{Can Large Language Models Detect Rumors on Social Media?}

\author{Qiang Liu$^{1,2}$ \qquad Xiang Tao$^{1,2}$ \qquad Junfei Wu$^{1,2}$ \qquad Shu Wu$^{1,2}$ \qquad Liang Wang$^{1,2}$ \\
        $^1$Center for Research on Intelligent Perception and Computing, \\ State Key Laboratory of Multimodal Artificial Intelligence Systems, \\ Institute of Automation, Chinese Academy of Sciences \\ 
        $^2$School of Artificial Intelligence, University of Chinese Academy of Sciences \\
 \texttt{qiang.liu@nlpr.ia.ac.cn,} \texttt{\{xiang.tao,junfei.wu\}@cripac.ia.ac.cn} \\  \texttt{\{shu.wu,wangliang\}@nlpr.ia.ac.cn}}

\begin{document}
\maketitle
\begin{abstract}
In this work, we investigate to use Large Language Models (LLMs) for rumor detection on social media.
However, it is challenging for LLMs to reason over the entire propagation information on social media, which contains news contents and numerous comments, due to LLMs may not concentrate on key clues in the complex propagation information, and have trouble in reasoning when facing massive and redundant information.
Accordingly, we propose an LLM-empowered Rumor Detection (LeRuD) approach, in which we design prompts to teach LLMs to reason over important clues in news and comments, and divide the entire propagation information into a Chain-of-Propagation for reducing LLMs' burden.
We conduct extensive experiments on the Twitter and Weibo datasets, and LeRuD outperforms several state-of-the-art rumor detection models by $3.2\%$ to $7.7\%$.
Meanwhile, by applying LLMs, LeRuD requires no data for training, and thus shows more promising rumor detection ability in few-shot or zero-shot scenarios.
\end{abstract}

\section{Introduction}

With the development of social media, users can access to information more easily, and messages can be spread more rapidly.
However, at the same time, rumors can also be spread to the public more easily and widely.
It is necessary to conduct research on automatic rumor detection on social media \cite{castillo2011information,islam2020deep}, which is still very challenging.
Recently, Large Language Models (LLMs) \cite{zhao2023survey} have achieved success and applied widely.
This brings great convenience to people.
However, people can also use LLMs to intentionally generate great amount of unidentified information \cite{vykopal2023disinformation,chen2024can}, which aggravate the propagation situation of rumor on social media.
Thus, the success of LLMs makes automatic rumor detection more challenging and urgent.

\begin{figure}[t]
  \centering
  \includegraphics[width=\linewidth]{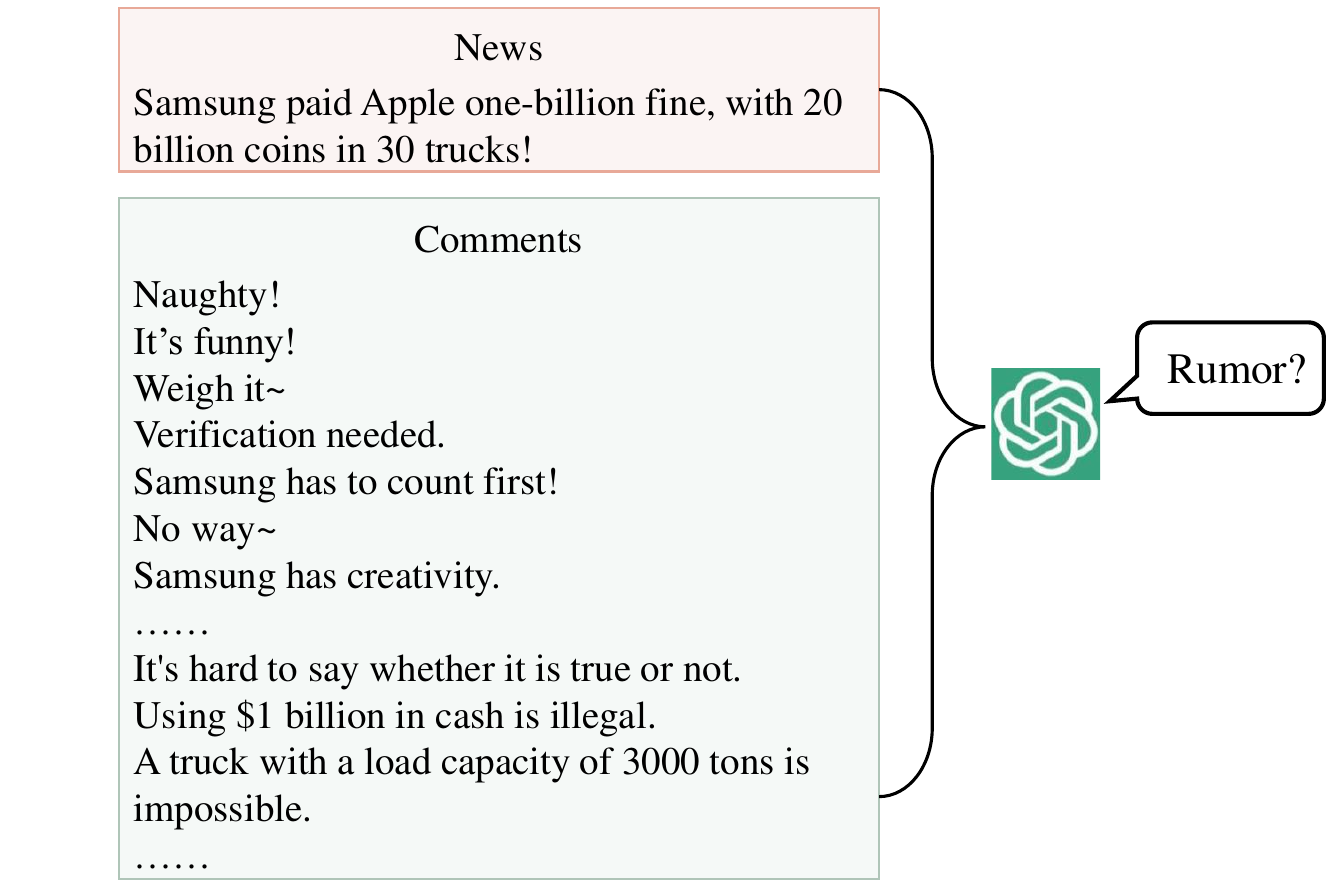}
  \caption{Can LLMs detect rumors on social media?}
\end{figure}

The key of rumor detection lies in the modeling of propagation patterns on social media, i.e., aggregating the features of original news and user comments.
Early approaches regards news and comments as sequences of texts, and applies recurrent networks \cite{ma2016detecting}, convolutional networks \cite{yu2017convolutional} or attention networks \cite{liu2018mining} for feature extraction.
Then, rumor detection models based on graph neural networks \cite{lu2020gcan,bian2020rumor} have been widely proposed and achieved state-of-the-art performances.

Recently, some research works attempt to apply LLMs in identifying the veracity of information \cite{chen2023combating}.
They mostly use LLMs to extract knowledge from texts \cite{yang2023rumor} or solely analyzing the news contents \cite{hu2024bad}.
The power of LLMs for analyzing the propagation information, which contains both news and comments, on social media has not been explored.
Meanwhile, as introduced by \cite{hu2024bad,chen2024can}, solely relying on textual features of news, it is hard to obtain promising detection results by LLMs.
This requires us to investigate the ability of LLMs for modeling the entire propagation information on social media.

However, using LLMs for modeling the propagation information is not easy.
Firstly, we find LLMs can not concentrate on the key clues in the propagation information, and thus may produce wrong predictions (Sec. \ref{sec:prompt1} and \ref{sec:prompt2}).
Secondly, the amount of comments is usually large, which brings great burden for LLMs to reason (Sec. \ref{sec:CoH}), as LLMs usually have input length limitations, and have trouble when facing long contexts or redundant information \cite{huang2023advancing,xie2023analysis}.

To teach LLMs to conduct reasoning over propagation information and overcome above shortcomings, we propose an \textbf{\underline{L}LM-\underline{e}mpowered \underline{Ru}mor \underline{D}etection} approach, called \textbf{LeRuD}.
Firstly, we draw experience from previous rumor detection models \cite{ma2016detecting,liu2018mining,ma2018rumor,yang2022reinforcement,jin2022towards,hu2024bad,przybyla2020capturing}, in which useful information for distinguish rumors has been highlighted.
We design proper prompts to teach LLMs to concentrate on the writing styles and commonsense mistakes of news, as well as rebuttals or conflicts in comments.
Secondly, we divide the propagation information into a Chain-of-Propagation, which enables LLMs to reason step-by-step in time order with a reasonable number of comments.

Then, to evaluate the effectiveness of the LeRuD approach on detecting rumors, we conduct experiments on the Twitter and Weibo datasets (Sec. \ref{sec:exps}).
We use the widely-applied GPT 3.5 \footnote{\url{https://chat.openai.com/}} in our implementation.
For smooth usage of GPT 3.5, we filter samples with ethics issues.
Meanwhile, to avoid data leakage, we also filter samples in which GPT 3.5 already knows about the facts in the news.
On the filtered datasets, LeRuD outperforms several state-of-the-art rumor detection models by $3.2\%$ to $7.7\%$.
To be noted, LeRuD achieves these results in zero-shot setting, while other models require training data.
In a word, \textbf{with our efforts, LLMs can well detect rumors on social media}.

Our main contributions can be listed as follows:
\begin{itemize}
    \item We propose to apply LLMs in detecting rumors on social media with the entire propagation information, which contains both news and comments.
    \item We design proper prompts to teach LLMs to concentrate on the key clues in news and comments, which enable LLMs to conduct reasoning from multiple views.
    \item We divide the propagation information of a piece of news into a Chain-of-Propagation, which enables LLMs to conduct reasoning step-by-step with reasonable amount of information in each time.
    \item We conduct extensive experiments, in which several state-of-the-art rumor detection models are compared. Results demonstrate the superiority of our LeRuD approach.
\end{itemize}

\begin{table*}[!ht]
	\centering
    \small
	\resizebox{\linewidth}{!}
	{\begin{tabular}{l|p{16cm}}
			\toprule
Vanilla Prompt &
There is a piece of news: ``\textcolor{blue}{Samsung paid Apple one-billion fine, with 20 billion coins in 30 trucks!}'' Verify the credibility of the news. Please choose the answer from the following options: A. Fake, B. Real.\\
			\midrule
Output &
I cannot verify the credibility of real-time or specific news events as my training only includes information up until January 2022, and I do not have the capability to access the internet for real-time updates.\\
\midrule
Rational Prompt &
There is a piece of news: ``\textcolor{blue}{Samsung paid Apple one-billion fine, with 20 billion coins in 30 trucks!}'' \textit{\textcolor{red}{Based on the writing style and the commonsense knowledge, estimate the credibility of the news.}} Based on above results, please choose the answer from the following options: A. Fake, B. Real.\\
\midrule
Output &
The news about Samsung paying Apple a one-billion fine with 20 billion coins in 30 trucks \textcolor{red}{\underline{seems highly implausible and}} \textcolor{red}{\underline{unrealistic}}. It lacks specific details, and the use of \textcolor{red}{\underline{such an unconventional method for payment}} raises doubts about the credibility of the information. Therefore, the most appropriate answer would be: \textbf{A. Fake}.\\
			\toprule
			\bottomrule
Vanilla Prompt &
There is a piece of news: ``\textcolor{blue}{Around 7 o'clock this morning, a major traffic accident occurred on the Yuncheng to Houma section of the Shanxi Dayun Expressway, 9.3 kilometers away from Houma. A total of 47 vehicles collided in a series, resulting in 3 deaths and 7 injuries.}'' Verify the credibility of the news. Please choose the answer from the following options: A. Fake, B. Real.\\
			\midrule
Output &
I cannot verify the credibility of real-time or specific news events as my training only includes information up until January 2022, and I do not have the capability to access the internet for real-time updates.\\
\midrule
Rational Prompt &
There is a piece of news: ``\textcolor{blue}{Around 7 o'clock this morning, a major traffic accident occurred on the Yuncheng to Houma section of the Shanxi Dayun Expressway, 9.3 kilometers away from Houma. A total of 47 vehicles collided in a series, resulting in 3 deaths and 7 injuries.}'' \textit{\textcolor{red}{Based on the writing style and the commonsense knowledge, estimate the credibility of the news.}} Based on above results, please choose the answer from the following options: A. Fake, B. Real.\\
\midrule
Output &
The provided news about a major traffic accident on the Yuncheng to Houma section of the Shanxi Dayun Expressway appears to be presented in a straightforward and factual manner. It includes \textcolor{red}{\underline{specific details such as the time of the incident, the location,}} \textcolor{red}{\underline{and the number of vehicles involved}}, as well as the \textcolor{red}{\underline{reported casualties (3 deaths and 7 injuries)}}. While I cannot independently verify the information, the news \textcolor{red}{\underline{seems plausible and is presented in a manner consistent with typical reporting}} of traffic accidents. Therefore, based on the writing style and common sense, it appears to be: \textbf{B. Real}.\\
			\bottomrule
	\end{tabular}}
        \caption{Analyzing news contents with GPT 3.5 on two samples. Without proper prompts, it is hard for the LLM to judge the credibility of the news. Thus, we design prompts (in \textit{\textcolor{red}{red italicized words}}) to teach the LLM to concentrate on the writing styles and commonsense mistakes, which enable the LLM to find important clues in writing details and conduct commonsense reasoning. Key clues during reasoning are highlighted with \textcolor{red}{\underline{red underlined words}}.}
	\label{tab:prompt1}
\end{table*}

\section{Related Work}

Rumor detection on social media is a major text mining task for automatically distinguishing untrustworthy information and maintaining the healthy development of the internet and society.
Extensive works have been done to detect rumors from different perspectives \cite{castillo2011information,islam2020deep}.

Early attempts on rumor detection mainly focus on extracting statistical features of the propagation process \cite{kwon2013prominent,ma2015detect}.
Then, deep neural networks for modeling sequential data have been widely applied in aggregating propagation information and detecting rumors on social media.
Among them, detection models based on recurrent neural networks \cite{ma2016detecting,ma2018rumor}, convolutional neural networks \cite{yu2017convolutional}, attentive networks \cite{liu2018mining} and generative
adversarial networks \cite{ma2019detect} have been extensively studied.
Meanwhile, their combinations have also been investigated for rumor detection \cite{liu2018early,yu2019attention}.
Recently, detection models based on graph neural networks \cite{bian2020rumor,lu2020gcan,nguyen2020fang,xu2022evidence} have achieved state-of-the-art performances.
Furthermore, some works investigate incorporate graph contrastive learning for better rumor detection performances \cite{sun2022rumor,wu2023adversarial}.
For better interpretability, reasoning over subgraph has been proposed \cite{jin2022towards, yang2022reinforcement}.
Meanwhile, for capturing temporal characteristics, dynamic graph has been constructed and applied in rumor detection \cite{sun2022ddgcn}.
Nowadays, considering practical applications, rumor detection in debiasing, few-shot or zero-shot scenarios have drawn much attention \cite{zhu2022generalizing,xu2023counterfactual,yue2023meta,lin2023zero,liu2023out}.

As LLMs have achieved great success, they also bring opportunities and challenges to misinformation identification \cite{chen2023combating}.
LLMs are able to easily generate misinformation \cite{vykopal2023disinformation,chen2024can}, which is hard to be distinguished by human or LLMs solely based on textual features \cite{chen2024can}.
CICAN \cite{yang2023rumor} uses GPT 3.5 as an auxiliary tool, for extracting entities from news and comments, and constructing relational graph.
ARG \cite{hu2024bad} uses LLMs for analyzing the news contents, and conclude that, LLMs are hard to beat fine-tuned small models on detecting fake news.
Thus, ARG incorporates insights generated by LLMs in small models for fake news detection.

\begin{table*}[!ht]
	\centering
    \small
	\resizebox{\linewidth}{!}
	{\begin{tabular}{l|p{16cm}}
			\toprule
Vanilla Prompt &
There is a piece of news: ``\textcolor{blue}{Around 7 o'clock this morning, a major traffic accident occurred on the Yuncheng to Houma section of the Shanxi Dayun Expressway, 9.3 kilometers away from Houma. A total of 47 vehicles collided in a series, resulting in 3 deaths and 7 injuries.}'' There are comments for the news: ``\textcolor{blue}{My god!}'' ``\textcolor{blue}{Blessing...}'' ``\textcolor{blue}{Why isn't the highway closed on such a foggy day?}'' ``\textcolor{blue}{Repost}'' ``\textcolor{blue}{Cherishing life!}'' ``\textcolor{blue}{How scary!}'' ``\textcolor{blue}{Blessing.}'' ``\textcolor{blue}{Drive carefully and pay attention to safety.}'' ``\textcolor{blue}{Shocking...}'' Based on the comments, verify the authenticity of the news. Please choose the answer from the following options: A. Fake, B. Real.
\\
			\midrule
Output &
Based on the provided comments, it's challenging to determine the authenticity of the news directly. The comments express shock, concern, and questioning regarding the incident, but they don't provide concrete evidence or information about the news source. Therefore, the answer is: \textbf{A. Fake}.\\
\midrule
Conflicting Prompt &
There is a piece of news: ``\textcolor{blue}{Around 7 o'clock this morning, a major traffic accident occurred on the Yuncheng to Houma section of the Shanxi Dayun Expressway, 9.3 kilometers away from Houma. A total of 47 vehicles collided in a series, resulting in 3 deaths and 7 injuries.}'' There are comments for the news: ``\textcolor{blue}{My god!}'' ``\textcolor{blue}{Blessing...}'' ``\textcolor{blue}{Why isn't the highway closed on such a foggy day?}'' ``\textcolor{blue}{Repost}'' ``\textcolor{blue}{Cherishing life!}'' ``\textcolor{blue}{How scary!}'' ``\textcolor{blue}{Blessing.}'' ``\textcolor{blue}{Drive carefully and pay attention to safety.}'' ``\textcolor{blue}{Shocking...}'' \textit{\textcolor{red}{Based on the comments, analyze whether there are any rebuttals or conflicts, and accordingly verify the authenticity of the news.}} Based on above results, please choose the answer from the following options: A. Fake, B. Real.
\\
\midrule
Output &
The comments mainly express shock, concern, and questions about the responsibility for the accident. Some comments criticize the decision not to close the expressway in thick fog, suggesting potential negligence. There is also a call for attention to safety in such weather conditions. There doesn't seem to be any direct rebuttal or conflict among the comments. Instead, \textcolor{red}{\underline{the comments} \underline{generally focus on expressing emotions, seeking information about the incident, and questioning the decision-making around} \underline{highway closure during foggy conditions.}} Based on the analysis of comments, \textcolor{red}{\underline{there is no clear evidence of conflicting}} \textcolor{red}{\underline{viewpoints that would cast doubt on the authenticity of the news.}} The concerns raised by commenters align with the details provided in the news report. Therefore, the answer is: \textbf{B. Real}.\\
			\bottomrule
	\end{tabular}}
        \caption{Analyzing comments with GPT 3.5 on a sample. Without proper prompts, the LLM tends to say the news is fake when there lacks clear supporting materials. However, in most cases, is it hard to have clear evidences, especially in early stage of propagation. Thus, we design prompts (in \textit{\textcolor{red}{red italicized words}}) to teach the LLM to concentrate on rebuttals or conflicts in comments, which lead the LLM to use crowd wisdom for judging the veracity of information. Key clues during reasoning are highlighted with \textcolor{red}{\underline{red underlined words}}.}
	\label{tab:prompt2}
\end{table*}

\section{Methodology}

\begin{figure*}[t]
  \centering
  \includegraphics[width=0.93\linewidth]{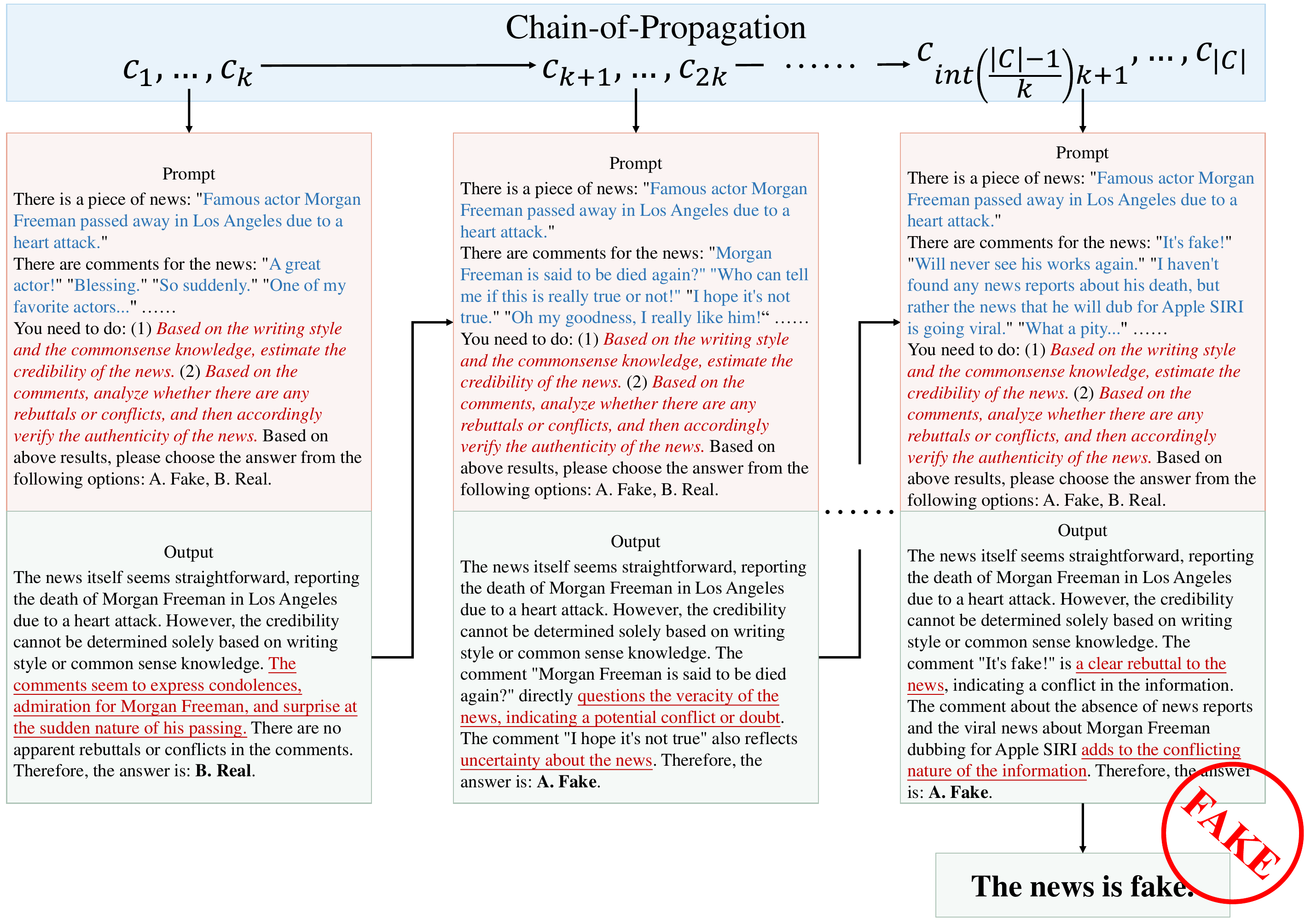}
  \caption{The procedure of the LeRuD approach. LeRuD divide the entire propagation information into a Chain-of-propagation with easy comment information for reasoning in each step. In each reasoning step, Rational and Conflicting Prompts, which are written in \textit{\textcolor{red}{red italicized words}}, are performed. Key findings during LeRuD's reasoning process are highlighted with \textcolor{red}{\underline{red underlined words}}.}
  \label{fig:model}
\end{figure*}

In this section, we detail our proposed LeRuD approach, which consists of prompts for reasoning over news and comments, as well as the Chain-of-Propagation reasoning process.
The whole procedure of LeRuD is demonstrated in Fig. \ref{fig:model}.

\subsection{Problem Formulation}

For a piece of news spreading on social media, we denote the news content as $n$, which is a paragraph of text.
During spreading, the news receives comments, denoted as $C = \{(c_1, t_1), (c_2, t_2), ...... \}$, in which $c_i$ denotes the textual content of the $i$-th comment, and $t_i$ means the relative posting time compared with the original news.
We aim to utilize LLMs for modeling both the news $n$ and comments $C$, and detecting rumors, i.e., identifying whether the news is fake or real.

\subsection{RP: Rational Prompts for News} \label{sec:prompt1}

We need design proper prompts to teach LLMs for analyzing news contents for rumor detection.
Here, we apply the widely-used GPT 3.5 as the LLM used in our proposed approach.
As shown in the two cases in Tab. \ref{tab:prompt1}, we try the vanilla prompts, which directly ask the LLM to infer the credibility of the news contents.
However, without proper prompts, it is hard for the LLM to make guesses about veracity.
The LLM does not know how to reason over the content that it has no knowledge about the described fact.
This is mainly because that, the LLM can not concentrate on some key clues in the news contents.

Accordingly to some previous study, writing styles of news are highlighted for misinformation identification \cite{przybyla2020capturing}, and commonsense reasoning is essential for distinguish fake news \cite{hu2024bad}.
Thus, we need to teach LLMs to reason from these aspect.
Accordingly, we design \textbf{Rational Prompts}, which ask the LLM to answer ``\textit{Based on the writing style and the commonsense knowledge, estimate the credibility of the news.}''
As shown in Tab. \ref{tab:prompt1}, with Rational Prompts, the LLM is able to find important clues in writing details and conduct commonsense reasoning, and give judgement on the veracity of news contents.
In the first case, the LLM correctly predicts the news is fake, by noticing that the event is not plausible from a commonsense perspective.
In the second case, the LLM correctly predicts the news is real, via noticing some accident details, and concluding that the writing style is consistent with typical reports.

\subsection{CP: Conflicting Prompts for Comments} \label{sec:prompt2}

Furthermore, we need to design prompts to teach LLMs to reason over comments in propagation.
As shown in Tab. \ref{tab:prompt2}, we try the vanilla prompts, which directly ask the LLM to verify the authenticity of news based on given comments.
However, with such prompts, the LLM tends to predict the news is fake when there lacks clear supporting materials.
In most cases, is it hard to have enough clear evidences, especially in the early stage of propagation.
This makes it easy for the LLM to mistakenly recognize real news as fake news.

Conflicts, rebuttals and skepticism in comments have been proven to be key features for distinguish misinformation \cite{ma2016detecting,liu2018mining,ma2018rumor,lu2020gcan,yu2017convolutional,yang2022reinforcement,jin2022towards}.
Considering the difficulty of having sufficient supporting materials, we design \textbf{Conflicting Prompts}, which ask the LLM to answer ``\textit{Based on the comments, analyze whether there are any rebuttals or conflicts, and accordingly verify the authenticity of the news.}''
As shown in Tab. \ref{tab:prompt2}, with Conflicting Prompts, the LLM correctly predicts the news is real, via noticing that people are all discussing about the event itself, and no conflicting information is posted.

\begin{figure}
	\centering
	\subfigure[Results on Twitter.]{
		\begin{minipage}[b]{0.225\textwidth}
			\includegraphics[width=1\textwidth]{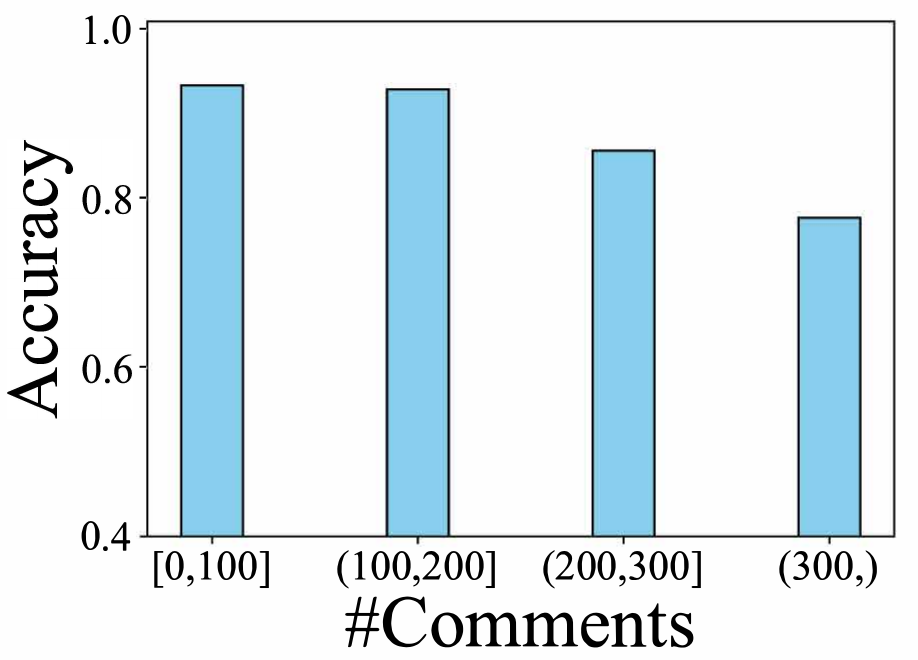}
		\end{minipage}
	}
	\subfigure[Results on Weibo.]{
		\begin{minipage}[b]{0.225\textwidth}
			\includegraphics[width=1\textwidth]{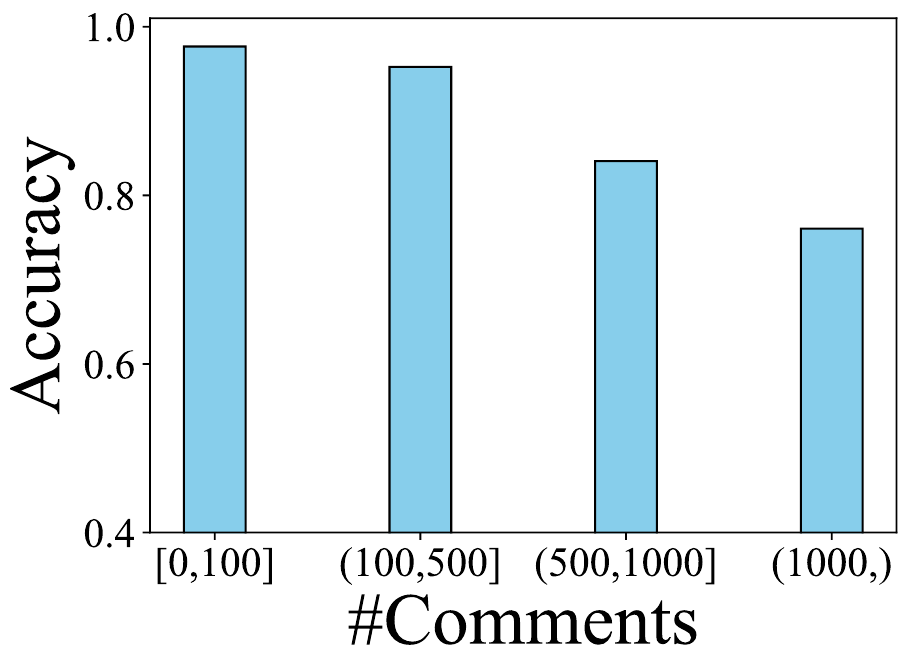}
		\end{minipage}
	}
	\caption{LLMs have trouble with too many comments.}
	\label{fig:amount}
\end{figure}

\subsection{CoP: Chain-of-Propagation} \label{sec:CoH}

Though Rational and Conflicting Prompts can teach LLMs to reason for rumor detection, LLMs still have trouble when facing too many comments.
In real social media scenarios, the number of comments is usually very large.
However, LLMs usually have limitations in input length.
And more seriously, LLMs can not well reason over long contexts or redundant information \cite{huang2023advancing,xie2023analysis}.
We test above Rational and Conflicting Prompts with GPT 3.5 on the Twitter and Weibo datasets (Sec. \ref{sec:data}), and results are shown in Fig. \ref{fig:amount}.
When the input length of a sample exceeds the limit, we randomly select part comments belonging to the sample as input.
It is clear that, when a sample has too many comments, the detection accuracy drops sharply, which means the LLM can not well reason over excessive information.

In recent research of LLMs, a complex task, which is hard for LLMs, is divided into a series of easy tasks for LLMs to execute \cite{wei2022chain,wang2023self,yao2023tree,besta2024graph}.
This enhances LLMs' ability for complex tasks.
Thus, we propose to divide the proposition information into a \textbf{Chain-of-Propagation (CoP)}, which enables LLMs to conduct reasoning more easily.
Considering comments in a time period usually share similar attitudes, we construct CoP in order of comments' posting time, and select $k$ comments in each reasoning step.
As shown in Fig. \ref{fig:amount}, the LLM can deal with up to $100$ comments easily, so we set $k=100$.
Reasoning steps in a CoP are run in the same session of LLM, so that latter reasoning steps can refer to the results of former reasoning steps.
Moreover, we use the result in the last step as the final estimation, considering the last step aggregates all the information.

\begin{table*}[htbp]
  \centering
  \caption{Results of performance comparison and ablation study.}
    \begin{tabular}{r|cccc|cccc}
    \toprule
    \multicolumn{1}{c|}{\multirow{2}[2]{*}{Approach}} & \multicolumn{4}{c|}{Twitter}  & \multicolumn{4}{c}{Weibo} \\
          & Accuracy & Precision & Recall & F1-score & Accuracy & Precision & Recall & F1-score \\
    \midrule
    \multicolumn{1}{c|}{BIGCN} & 0.8882  & 0.8739  & 0.8940  & 0.8838  & 0.9258  & 0.9133  & 0.9335  & 0.9233  \\
    \multicolumn{1}{c|}{DDGCN} & 0.8982  & 0.8834  & 0.9078  & 0.8954  & 0.9414  & 0.9390  & 0.9483  & 0.9436  \\
    \multicolumn{1}{c|}{GACL} & 0.9042  & 0.8884  & 0.9171  & 0.9025  & 0.9354  & 0.9294  & 0.9409  & 0.9351  \\
    \multicolumn{1}{c|}{CICAN} & 0.9082  & 0.8855  & 0.9263  & 0.9054  & 0.9390  & 0.9298  & 0.9458  & 0.9377  \\
    \multicolumn{1}{c|}{ARG} & 0.8902  & 0.8628  & 0.8986  & 0.8803  & 0.9043  & 0.9022  & 0.9089  & 0.9055  \\
    \midrule
    \multicolumn{1}{c|}{LeRuD} & \textbf{0.9401}  & \textbf{0.9156}  & 0.9493  & \textbf{0.9321}  & \textbf{0.9809}  & \textbf{0.9732}  & 0.9852  & \textbf{0.9792}  \\
    w/o RP   & 0.9261  & 0.9107  & 0.9401  & 0.9252  & 0.9689  & 0.9635  & 0.9754  & 0.9694  \\
    w/o CP   & 0.9142  & 0.8525  & \textbf{0.9585}  & 0.9024  & 0.9342  & 0.8777  & \textbf{0.9901}  & 0.9306  \\
    w/o CoP  & 0.9281  & 0.9148  & 0.9401  & 0.9273  & 0.9330  & 0.9249  & 0.9409  & 0.9328  \\
    \bottomrule
    \end{tabular}%
  \label{tab:results}%
\end{table*}%

\section{Experiments} \label{sec:exps}

In this section, we conduct extensive experiments, to verify whether LLMs can detect rumors on social media with our efforts.

\subsection{Data Preparation} \label{sec:data}

Firstly, we introduce the data preparation process, which enables us to conduct sufficient evaluation.

\textbf{Datasets}: For evaluation, we combine Twitter15 and Twitter16 as the \textbf{Twitter} dataset \cite{ma2017detect}, and also incorporate the \textbf{Weibo} dataset \cite{ma2018rumor}.
They are collected from two popular social media platforms, i.e., Twitter and Weibo.
The average numbers of comments of each news in Twitter and Weibo are $232$ and $816$ respectively.
We only use the verified false rumors (fake news) and non-rumors (real news) in Twitter.

\textbf{Data Filtering}: For smooth and low-cost usage of GPT 3.5, we filter samples with \textbf{ethics issues} (such as violence, pornography and discrimination) and \textbf{repeated news contents}.
We also filter samples containing mainly \textbf{subjective expressions}, for they are not factually real or fake.
Meanwhile, for fair comparison, we need to filter samples with \textbf{data leakage risks} raised by the usage of LLMs, in which GPT 3.5 already exactly knows about the facts in the news.
Thus, we direly ask GPT 3.5 to verify the veracity of each news with the Vanilla Prompts in Tab. \ref{tab:prompt1}, and only keep samples in which GPT 3.5 has no idea about the facts and can not give answers.
In contrast, when GPT 3.5 knows about the fact in a piece of news, it will detail the fact and clearly judge the veracity.

\subsection{Settings}

The, we detail the settings of our experiments.

\textbf{Baselines}: Firstly, we compare with some state-of-the-art rumor detection models based graph neural networks: \textbf{BIGCN} \cite{bian2020rumor}, \textbf{DDGCN} \cite{sun2022ddgcn} and \textbf{GACL} \cite{sun2022rumor}.
Secondly, we also compare with two rumor detection models with LLMs: \textbf{CICAN} \cite{yang2023rumor} and \textbf{ARG} \cite{hu2024bad}.

\textbf{Metrics}: We use \textbf{Accuracy}, \textbf{Precision}, \textbf{Recall} and \textbf{F1-score} for performance evaluation.
Precision, Recall and F1-score are calculated according to the rumor detection results.

\textbf{Zero-shot Setting}: LeRuD is performed in zero-shot setting, in which no training data is needed, and the evaluation is done with all samples in the dataset.
We use GPT 3.5 as the LLM in LeRuD.

\textbf{$5$-fold Cross-validation Setting}: For other compared baselines, we perform $5$-fold cross-validation.
The whole process is run for $5$ times, and in each time, $3$ folds of samples are used for training, $1$ fold of samples are used for validation, and $1$ fold of samples are used for testing.
After 5-time running, all samples have been used for testing, which stays the same as in the zero-shot setting.

\begin{table*}[!ht]
	\centering\small
	\resizebox{\linewidth}{!}
	{\begin{tabular}{l|p{16cm}}
			\toprule
Prompt &
There is a piece of news: ``\textcolor{blue}{Samsung paid Apple one-billion fine, with 20 billion coins in 30 trucks!}'' Verify the credibility of the news. Please choose the answer from the following options: A. Fake, B. Real.\\
			\midrule
Output &
I cannot verify the credibility of real-time or specific news events as my training only includes information up until January 2022, and I do not have the capability to access the internet for real-time updates.\\
			\bottomrule
	\end{tabular}}
        \caption{When asked to estimate the veracity of Sample ``3484403570896812'' in the Weibo dataset, GPT 3.5 has no idea about the fact.}
	\label{tab:exampleA}
\end{table*}

\begin{table*}[!ht]
	\centering\small
	\resizebox{\linewidth}{!}
	{\begin{tabular}{l|p{16cm}}
			\toprule
Prompt &
There is a piece of news: ``\textcolor{blue}{Ex-New York Giants safety Tyler Sash, who died at 27, found to have degenerative brain disease.}'' Verify the credibility of the news. Please choose the answer from the following options: A. Fake, B. Real.\\
			\midrule
Output &
As of my last update in January 2022, the information you provided was accurate. Tyler Sash, a former New York Giants safety, passed away at the age of 27 and was found to have chronic traumatic encephalopathy (CTE), a degenerative brain disease often associated with head trauma in contact sports. Therefore, the answer is: \textbf{B. Real}.\\
			\bottomrule
	\end{tabular}}
        \caption{When asked to estimate the veracity of sample ``692142338890661888'' in the Twitter dataset, GPT 3.5 provides details about the fact, and confirms the news is real.}
	\label{tab:exampleB}
\end{table*}

\subsection{Performance Comparison}

In Tab. \ref{tab:results}, we compare LeRuD with several state-of-the-art rumor detection models.
Among the three graph-based detection models, DDGCN and GACL obviously outperform BIGCN.
With the help of GPT 3.5 for extracting entities, CICAN achieves similar or better performances compared to the graph-based models.
ARG considers only news contents, so it underperforms other detection models which capture both news and comments.
However, with insights generated by LLMs, ARG still achieves high accuracy of about $90\%$.
Meanwhile, our proposed LeRuD approach achieves the best performances on both dataset evaluated by all metrics.
Specifically, LeRuD outperforms the best baselines on Twitter and Weibo by $3.19\%$ and $3.95\%$ respectively.
These results strongly demonstrate the superiority of LeRuD.
Moreover, \textbf{LeRuD requires no data for training}, so it can consistently achieve promising rumor detection performances in few-shot or zero-shot scenarios, and thus is more suitable for real-world rumor detection.

\subsection{Ablation Study}

In Tab. \ref{tab:results}, we also demonstrate the ablation study of LeRuD, including three variants: LeRuD w/o RP, LeRuD w/o CP and LeRuD w/o CoP.
\textbf{LeRuD w/o RP} denotes LeRuD without Rational Prompts.
It decreases the performances of LeRuD by approximately $1\%$, but still outperforms other baselines.
\textbf{LeRuD w/o CP} denotes LeRuD without Conflicting Prompts.
The declines of accuracy and f1-score are $2.5\%$ to $4.8\%$. 
LeRuD w/o CP achieves the highest recall values, but also achieves the lowest precision values.
This is because that, without Conflicting Prompts, LLMs may mistakenly classify real news as fake news.
\textbf{LeRuD w/o CoP} Denotes LeRuD without Chain-of-Propagation.
The declines of accuracy and f1-score are about $1\%$ and $4.5\%$ on Twitter and Weibo respectively.
The performance declines on Weibo are much larger, because the average number of comments in the Weibo dataset is much larger, according to the statistics in Sec. \ref{sec:data}.
This strongly proves that, LLMs have trouble in reasoning on excessive information.
Above observations demonstrate the necessity of the three modules proposed in LeRuD.

\begin{figure}[t]
	\centering
	\subfigure[Results on Twitter.]{
		\begin{minipage}[b]{0.225\textwidth}
			\includegraphics[width=1\textwidth]{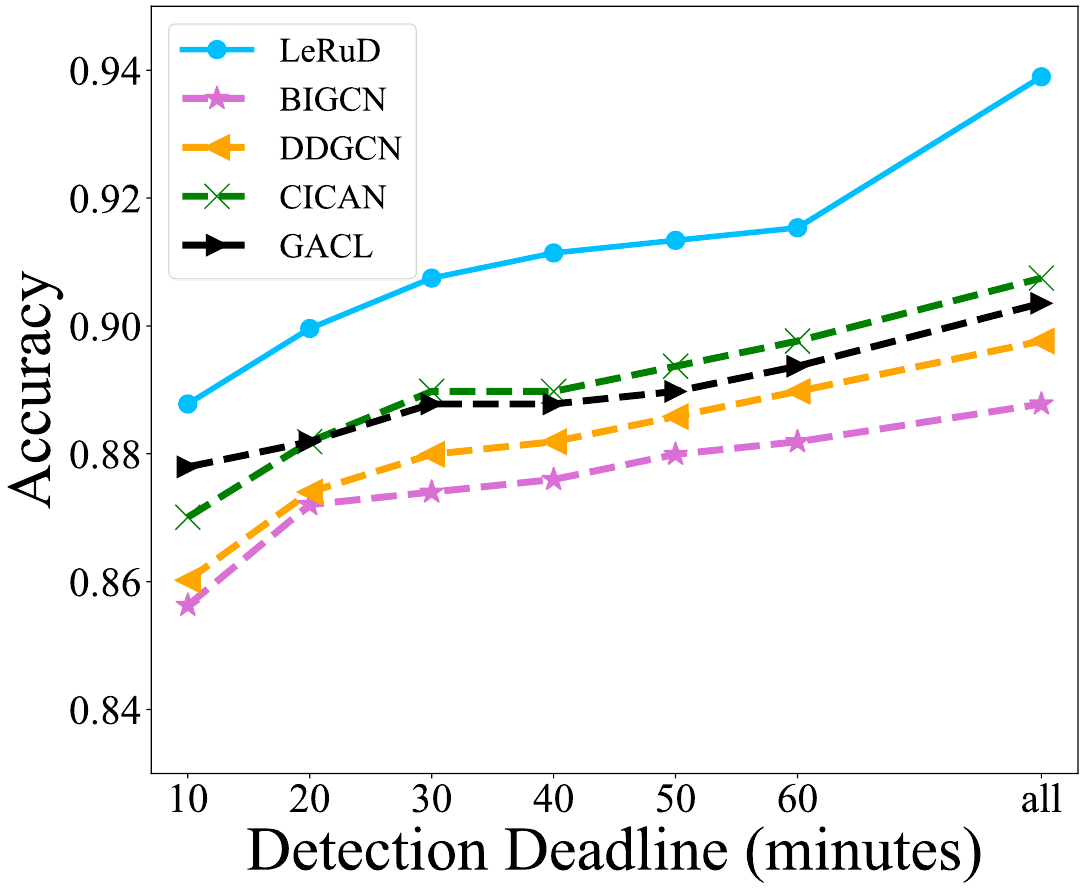}
		\end{minipage}
	}
	\subfigure[Results on Weibo.]{
		\begin{minipage}[b]{0.225\textwidth}
			\includegraphics[width=1\textwidth]{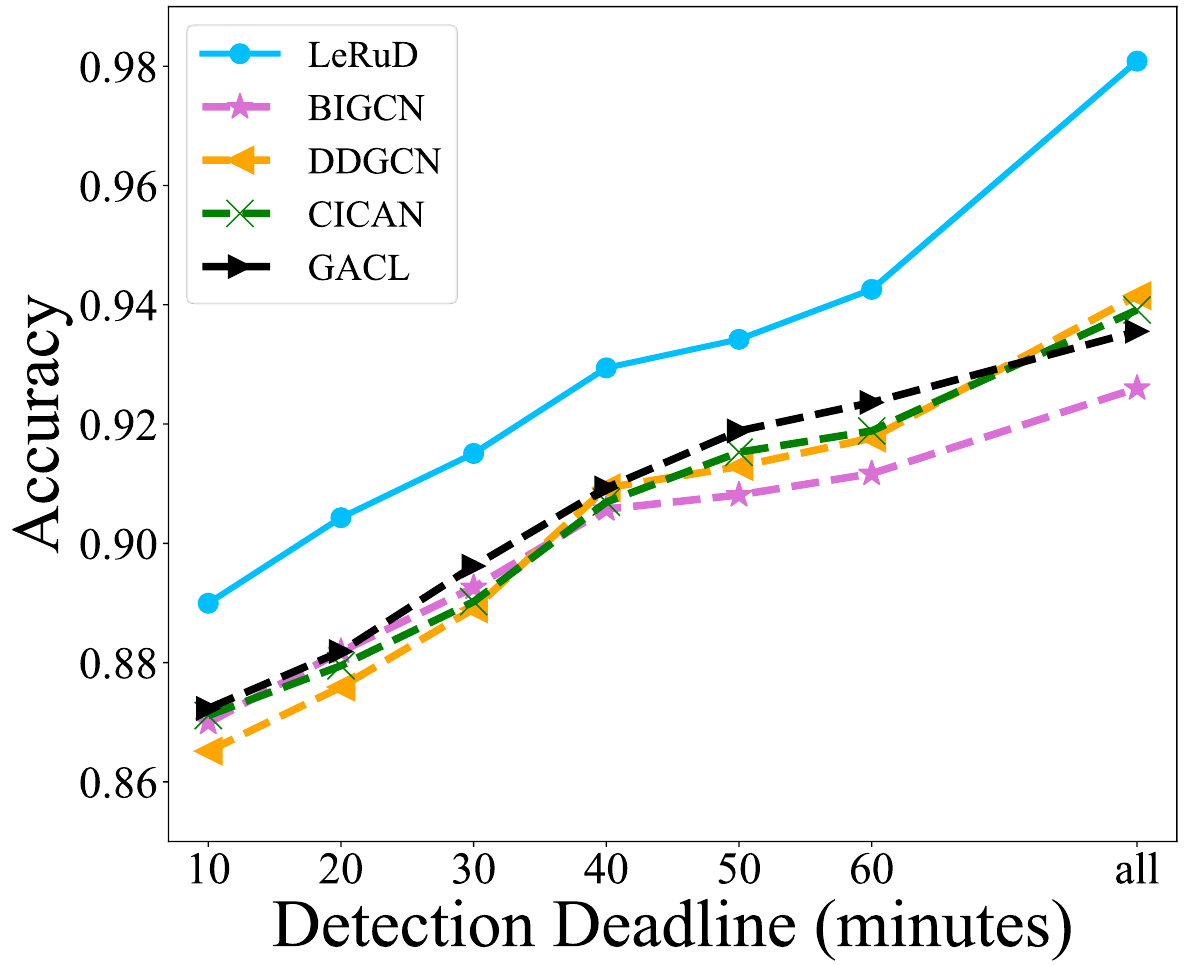}
		\end{minipage}
	}
	\caption{Early rumor detection results evaluated by accuracy with different detection deadlines.}
	\label{fig:early}
\end{figure}

\subsection{Early Detection}

As shown in Fig. \ref{fig:early}, we conduct early rumor detection with different detection deadlines.
ARG is not included, for it does not model the propagation information.
LeRuD constantly and significantly outperforms all the baselines with all the detection deadlines.
This shows that, LeRuD can perform well during early stage of propagation, and does not rely on the sufficiency of comment information.

\begin{figure}[t]
	\centering
	\subfigure[Results on Twitter.]{
		\begin{minipage}[b]{0.225\textwidth}
			\includegraphics[width=1\textwidth]{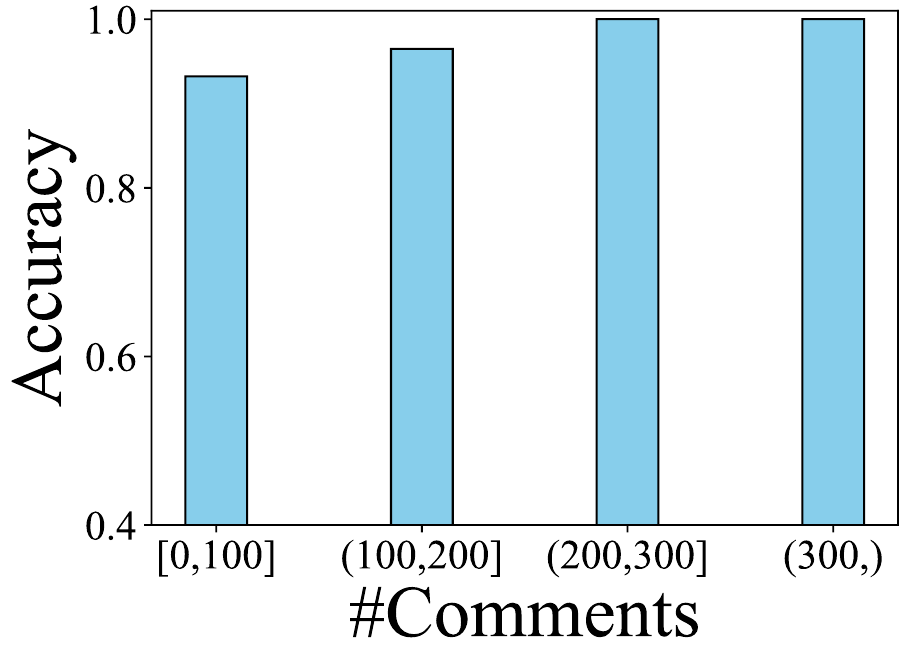}
		\end{minipage}
	}
	\subfigure[Results on Weibo.]{
		\begin{minipage}[b]{0.225\textwidth}
			\includegraphics[width=1\textwidth]{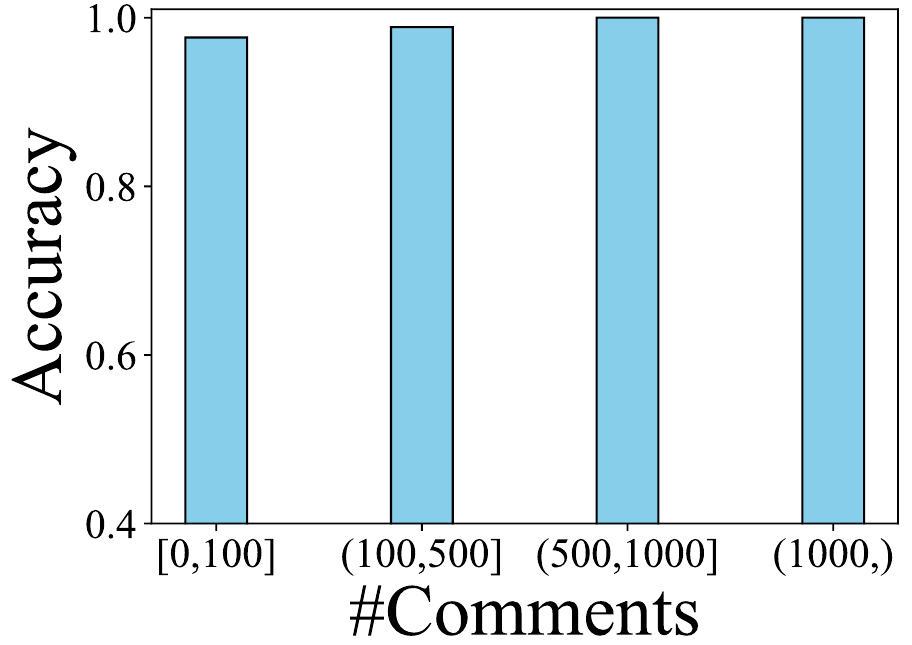}
		\end{minipage}
	}
	\caption{Accuracy of LeRuD under different numbers of comments. More comments, better performances.}
	\label{fig:amount_ours}
\end{figure}

\subsection{Impact of Comment Amount}

In Fig. \ref{fig:amount_ours}, we illustrate the accuracy of LeRuD under different numbers of comments.
With the help of CoP, LeRuD has no more trouble in conducting reasoning when facing great amount of comments.
Meanwhile, with more comments, LeRuD has even higher rumor detection accuracy, as there is more information available for analysis.
Above observations are totally different from those in Fig. \ref{fig:amount}.
In a word, dividing the propagation information into Chain-of-Propagation helps LLMs to conduct reasoning more easily.

\begin{table*}[!ht]
	\centering
    \small
	\resizebox{\linewidth}{!}
	{\begin{tabular}{l|p{16cm}}
			\toprule
Prompt of Step 1 &
There is a piece of news: ``\textcolor{blue}{Samsung paid Apple one-billion fine, with 20 billion coins in 30 trucks!}'' There are comments for the news: ``\textcolor{blue}{Naughty!}'' ``\textcolor{blue}{It's funny!}'' ``\textcolor{blue}{Weigh it~}'' ``\textcolor{blue}{Samsung has to count first!}'' ``\textcolor{blue}{How creativity!}'' ``\textcolor{blue}{Verification needed.}'' …… You need to do: (1) Based on the writing style and the commonsense knowledge, estimate the credibility of the news. (2) Based on the comments, analyze whether there are any rebuttals or conflicts, and then accordingly verify the authenticity of the news. Based on above results, please choose the answer from the following options: A. Fake, B. Real.\\
			\midrule
Output of Step 1 &
Based on the writing style and commonsense knowledge, the news seems highly unlikely and lacks credibility. Paying a fine with 20 billion coins in 30 trucks \textcolor{red}{\underline{sounds more like a creative or satirical piece rather than a factual event}}. Additionally, such a method of payment \textcolor{red}{\underline{raises practical concerns and seems impractical}}. Analyzing the comments, we can observe \textcolor{red}{\underline{a mix}} \textcolor{red}{\underline{of skepticism, humor, and requests for verification}}. Comments like ``Naughty!,'' ``Weigh it,'' and ``Samsung has to count first!'' \textcolor{red}{\underline{suggest doubts and a need for verification}}. The comment ``How creativity!'' acknowledges the imaginative aspect, and ``Verification needed'' explicitly \textcolor{red}{\underline{calls for confirmation}}. Therefore, the answer is: \textbf{A. Fake}.\\
\midrule
Prompt of Step 2 &
There is a piece of news: ``\textcolor{blue}{Samsung paid Apple one-billion fine, with 20 billion coins in 30 trucks!}'' There are comments for the news: ``\textcolor{blue}{It's hard to say whether it is true or not.}'' ``\textcolor{blue}{Using \$1 billion in cash is illegal.}'' ``\textcolor{blue}{Repost}'' ``\textcolor{blue}{A truck with a load capacity of 3000 tons is impossible.}'' ``\textcolor{blue}{Well done!}'' ......  You need to do: (1) Based on the writing style and the commonsense knowledge, estimate the credibility of the news. (2) Based on the comments, analyze whether there are any rebuttals or conflicts, and then accordingly verify the authenticity of the news. Based on above results, please choose the answer from the following options: A. Fake, B. Real.\\
			\midrule
Output of Step 2 &
Based on the writing style and commonsense knowledge, the news still appears unlikely and lacks credibility. Paying a one-billion dollar fine with 20 billion coins in 30 trucks \textcolor{red}{\underline{seems impractical and raises legal and logistical issues}}. Analyzing the comments, there are several expressions of doubt and skepticism. Comments like ``It's hard to say whether it is true or not,'' ``Using \$1 billion in cash is illegal,'' and ``A truck with a load capacity of 3000 tons is impossible'' \textcolor{red}{\underline{highlight concerns about the feasibility}} \textcolor{red}{\underline{and legality}} of the described scenario. The comment ``Well done!'' seems to be sarcastic or ironic, suggesting a lack of belief in the news. Therefore, the answer is: \textbf{A. Fake}.\\
			\midrule
Final Output &
The final estimation of the veracity of the news is: \textbf{The news is fake.}\\
			\bottomrule
	\end{tabular}}
        \caption{Prompts and outputs of performing the LeRuD approach on sample ``3484403570896812'' in the Weibo dataset. Key clues during LeRuD's reasoning process are highlighted with \textcolor{red}{\underline{red underlined words}}. LeRuD correctly identifies the sample as fake, and the explanations are pretty reasonable from the commonsense perspective.}
	\label{tab:example}
\end{table*}

\subsection{The Data Leakage Risk}

As in Sec. \ref{sec:data}, we filter samples with data leakage risks, in which GPT 3.5 already exactly knows about the facts in the news.
Examples with and without data leakage risks are shown in Tab. \ref{tab:exampleA} and \ref{tab:exampleB} respectively.
More examples can be found in App. \ref{sec:ex1}.
According to these examples, \textbf{we can clearly observe totally different outputs by the LLM between samples with and without data leakage risks}.
This makes us easy to distinguish samples with and without data leakage risks.
That is to say, after filtering these samples, we can avoid the data leakage risk, and experimental comparison can be fairly done among LeRuD and the baselines.

\subsection{Explanations}

We can further refer to the explanations generated by the LLM, to ensure that LeRuD does not take use of the exact factual knowledge for detecting rumors.
An example is shown in Tab. \ref{tab:example}, and more examples can be found in App. \ref{sec:ex2}.
According to these explanations, we can summarize the key clues in the reasoning process of LeRuD as follows:

\textbf{The regularity of writing}:
When the writing is informal, or the tone seems suspicions, LeRuD tends to judge the samples as fake (Tab. \ref{tab:EX-676120162018451456} and \ref{tab:EX-3507985437364629}).
When the writing is formal, or consistent with typical reporting styles, LeRuD tends to judge the samples as real (Tab. \ref{tab:EX-3909084371985775} and \ref{tab:EX-3911539469801102}-\ref{tab:EX-692764799315398656}).

\textbf{The adequacy of details}:
When there lacks enough details of events in the news, LeRuD tends to judge the samples as fake (Tab. \ref{tab:EX-3498854609883720}, \ref{tab:EX-676120162018451456}, \ref{tab:EX-538412662858588160} and \ref{tab:EX-3507985437364629}).
When details are rich, LeRuD tends to judge the samples as real (Tab. \ref{tab:EX-3911539469801102}, \ref{tab:EX-764497123530375169} and \ref{tab:EX-767203096472719364}).

\textbf{The plausibility of contents}:
When the contents seem impractical, or there are conflicts within the contents, or the events do not conform to commonsense logic, LeRuD tends to judge the samples as fake (Tab. \ref{tab:example} and \ref{tab:EX-3507985437364629}).
When the contents seem plausible and rational from the commonsense perspective, or the events have large chances to occur in certain circumstances, LeRuD tends to judge the samples as real (Tab. \ref{tab:EX-3911539469801102} - \ref{tab:EX-693119705469587456}).

\textbf{The attitudes of public}:
When the attitudes in the comments contains massive skepticism, questioning and rebuttals, or even some comments directly claim the news is fake, LeRuD tends to judge the samples as fake (Tab. \ref{tab:example} and \ref{tab:EX-3498854609883720} - \ref{tab:EX-3507985437364629}).
When comments seldom express skepticism or rebuttals, and people are discussing and commenting on the news events themselves, LeRuD tends to judge the samples as real (Tab. \ref{tab:EX-3909084371985775} - \ref{tab:EX-3914951724449159}).

\textbf{The consistency of comments}:
When details in the comments conflict with the news events, or demonstrates the irrationality of the news, LeRuD tends to judge the samples as fake (Tab. \ref{tab:example}, \ref{tab:EX-3498854609883720}, \ref{tab:EX-3465261540071546} and \ref{tab:EX-504448356584677377}).
When details in the comments conform to the news, LeRuD tends to judge the samples as real (Tab. \ref{tab:EX-3909084371985775} and \ref{tab:EX-3912024620676243}).

In summary, LeRuD detects rumor on social media via reasoning from commonsense perspectives, and \textbf{does not take usage of the LLM's exact prior knowledge about news facts}.

\section{Conclusion}
In this paper, we investigate to use LLMs for rumor detection on social media, and propose a novel approach called LeRuD.
LeRuD designs proper prompts to teach LLMs to reason over important clues in news and comments.
Meanwhile, we divide the entire propagation information of a piece of news into a Chain-of-Propagation for leading LLMs to conduct reasoning step-by-step with reasonable amount of information in each time.
Then, we implement LeRud with the widely-used GPT 3.5, and conduct extensive experiments on the Twitter and Weibo datasets.
LeRud can outperform several state-of-the-art rumor detection models with large margins.
In summary, with our efforts, LLMs can well detect rumors on social media.


\clearpage
\section*{Limitations}
In this paper, though we have taught LLMs to detect rumors on social media, our work still has limitations in the following two aspects.
Firstly, running LLMs for rumor detection requires relatively high costs, such as the usage of high-performance GPUs, or the calls of APIs, or inputs of web interfaces.
Secondly, graph neural networks have great ability for modeling structural data, while structural data is hard to be well described in texts.
Thus, using LLMs for analyzing complex propagation structures, or combining the abilities of LLMs and graph neural networks, has not been explored.
In the future, we shall put more efforts to overcome above limitations.

\bibliography{ref}

\appendix

\begin{table}[!htbp]
  \centering
  \caption{The ratio of samples with data leakage risks, in which GPT 3.5 already knows about the facts in the news, on the Twitter and Weibo datasets.}
    \begin{tabular}{c|cc}
    \toprule
          & Twitter & Weibo \\
    \midrule
    Ratio of Data Leakage & 0.1880 & 0.0434 \\
    \bottomrule
    \end{tabular}%
  \label{tab:leak}%
\end{table}%

\section{Examples about Data Leakage Risks}
\label{sec:ex1}

In Tab. \ref{tab:leak}, we calculate the ratio of samples with data leakage risks, i.e., samples to be filtered.
The ratio on the Twitter dataset is much larger than that in the Weibo dataset.
This may be caused by the data sources used for training GPT 3.5.
News in English has been widely collected, while news in Chinese has been seldom used for the training process of GPT 3.5.

Moreover, we further illustrate several samples, some of which GPT 3.5 exactly knows about the facts, some of which it does not know.
In Tab. \ref{tab:exA01}-\ref{tab:exA03}, there are samples in which GPT 3.5 has no idea about the facts, and the LLM can not give answers on the veracity of news.
In Tab. \ref{tab:exB01}-\ref{tab:exB03}, the LLM provides verified descriptions about the facts, and confirms they are fake news.
In Tab. \ref{tab:exC01}-\ref{tab:exC03}, the LLM provides more details about the facts, and confirms they are real news.

Accordingly to these examples, there are mainly two types of samples with data leakage risks: (1) real time news which has been record in the data for training GPT 3.5; (2) non real-time general knowledge which mainly concentrate in the fields such as health and history.
In actual online rumor detection systems, non real-time knowledge (type 2) can actually be helpful to the detection process, while real-time knowledge (type 1) is hard to be remembered by LLMs timely.

\section{Examples of LeRuD}
\label{sec:ex2}

In this section, we provide more examples of LeRuD's running results in Tab. \ref{tab:EX-3498854609883720} - \ref{tab:EX-3914951724449159}.

\begin{table*}[!ht]
	\centering
	\resizebox{\linewidth}{!}
	{\begin{tabular}{l|p{16cm}}
			\toprule
Prompt &
There is a piece of news: ``\textcolor{blue}{Famous actor Morgan Freeman passed away in Los Angeles due to a heart attack.}'' Verify the credibility of the news. Please choose the answer from the following options: A. Fake, B. Real.\\
			\midrule
Output &
I cannot verify the credibility of real-time or specific news events as my training only includes information up until January 2022, and I do not have the capability to access the internet for real-time updates.\\
			\bottomrule
	\end{tabular}}
        \caption{Sample ``3547977337564354'' in the Weibo dataset. GPT 3.5 has no idea about the fact.}
	\label{tab:exA01}
\end{table*}

\begin{table*}[!ht]
	\centering
	\resizebox{\linewidth}{!}
	{\begin{tabular}{l|p{16cm}}
			\toprule
Prompt &
There is a piece of news: ``\textcolor{blue}{Top North Korean diplomat who disappeared from London embassy has defected to South Korea, Seoul.}'' Verify the credibility of the news. Please choose the answer from the following options: A. Fake, B. Real.\\
			\midrule
Output &
I cannot verify the credibility of real-time or specific news events as my training only includes information up until January 2022, and I do not have the capability to access the internet for real-time updates.\\
			\bottomrule
	\end{tabular}}
        \caption{Sample ``765859710503378944'' in the Twitter dataset. GPT 3.5 has no idea about the fact.}
	\label{tab:exA02}
\end{table*}

\begin{table*}[!ht]
	\centering
	\resizebox{\linewidth}{!}
	{\begin{tabular}{l|p{16cm}}
			\toprule
Prompt &
There is a piece of news: ``\textcolor{blue}{ALERT: Tsunami advisory issued for Southern California from 3-7 a.m. local time. Stay away from ocean and shoreline.}'' Verify the credibility of the news. Please choose the answer from the following options: A. Fake, B. Real.\\
			\midrule
Output &
I cannot verify the credibility of real-time or specific news events as my training only includes information up until January 2022, and I do not have the capability to access the internet for real-time updates.\\
			\bottomrule
	\end{tabular}}
        \caption{Sample ``765859710503378944'' in the Twitter dataset. GPT 3.5 has no idea about the fact.}
	\label{tab:exA03}
\end{table*}

\begin{table*}[!ht]
	\centering
	\resizebox{\linewidth}{!}
	{\begin{tabular}{l|p{16cm}}
			\toprule
Prompt &
There is a piece of news: ``\textcolor{blue}{GMO foods weaken reproductive ability. Eating them mean chronic suicide.}'' Verify the credibility of the news. Please choose the answer from the following options: A. Fake, B. Real.\\
			\midrule
Output &
As of my last knowledge update in January 2022, there is no widespread scientific consensus that GMO (Genetically Modified Organism) foods weaken reproductive ability or lead to chronic suicide. Numerous studies have been conducted on the safety of GMOs, and the majority of scientific organizations, including the World Health Organization (WHO) and the National Academy of Sciences, have stated that GMOs currently approved for sale are safe to eat. Therefore, the answer is: \textbf{A. Fake}.\\
			\bottomrule
	\end{tabular}}
        \caption{Sample ``8535326047'' in the Weibo dataset. GPT 3.5 confirms the news is fake.}
	\label{tab:exB01}
\end{table*}

\begin{table*}[!ht]
	\centering
	\resizebox{\linewidth}{!}
	{\begin{tabular}{l|p{16cm}}
			\toprule
Prompt &
There is a piece of news: ``\textcolor{blue}{The screen car god Paul Walker was not dead and has been rescued.}'' Verify the credibility of the news. Please choose the answer from the following options: A. Fake, B. Real.\\
			\midrule
Output &
As of my last knowledge update in January 2022, Paul Walker, the actor known for his role in the "Fast and Furious" film series, passed away in a car accident on November 30, 2013. Any news claiming that he has been rescued and is not dead would be false. Therefore, the answer is: \textbf{A. Fake}.\\
			\bottomrule
	\end{tabular}}
        \caption{Sample ``3651171602494589'' in the Weibo dataset. GPT 3.5 confirms the news is fake.}
	\label{tab:exB02}
\end{table*}

\clearpage

\begin{table*}[!ht]
	\centering
	\resizebox{\linewidth}{!}
	{\begin{tabular}{l|p{16cm}}
			\toprule
Prompt &
There is a piece of news: ``\textcolor{blue}{NASA Confirms Earth Will Experience 15 Days Of Complete Darkness.}'' Verify the credibility of the news. Please choose the answer from the following options: A. Fake, B. Real.\\
			\midrule
Output &
The news stating "NASA Confirms Earth Will Experience 15 Days Of Complete Darkness" is fake. There is no credible information or scientific basis to support such a claim. If this news were true, it would have had widespread and legitimate coverage from reputable sources. Therefore, the answer is: \textbf{A. Fake}.\\
			\bottomrule
	\end{tabular}}
        \caption{Sample ``620916279608651776'' in the Twitter dataset. GPT 3.5 confirms the news is fake.}
	\label{tab:exB03}
\end{table*}

\begin{table*}[!ht]
	\centering
	\resizebox{\linewidth}{!}
	{\begin{tabular}{l|p{16cm}}
			\toprule
Prompt &
There is a piece of news: ``\textcolor{blue}{The Lushun Massacre, an unforgettable Japanese crime! After the Japanese army captured the fortress of Lushun, they carried out inhumane robbery, massacre, and rape inside the city. Over four days and three nights, approximately 20000 innocent civilians were brutally killed by the Japanese army's knives.}'' Verify the credibility of the news. Please choose the answer from the following options: A. Fake, B. Real.\\
			\midrule
Output &
The Lushun Massacre did occur, and the information in the news is generally accurate. Lushun, also known as Port Arthur, was a Chinese fortress that was captured by the Japanese army during the First Sino-Japanese War in 1894. There were reports of inhumane acts, including robbery, massacre, and rape, during the Japanese occupation of the city. The details may vary, and historical accounts might differ, but the basic premise of the massacre is real. Therefore, the answer is: \textbf{B. Real}.\\
			\bottomrule
	\end{tabular}}
        \caption{Sample ``3911526290906659'' in the Weibo dataset. GPT 3.5 confirms the news is real.}
	\label{tab:exC01}
\end{table*}

\begin{table*}[!ht]
	\centering
	\resizebox{\linewidth}{!}
	{\begin{tabular}{l|p{16cm}}
			\toprule
Prompt &
There is a piece of news: ``\textcolor{blue}{Colombian government and left-wing Farc rebel movement agree ceasefire to end decades-long conflict.}'' Verify the credibility of the news. Please choose the answer from the following options: A. Fake, B. Real.\\
			\midrule
Output &
As of my last knowledge update in January 2022, a historic peace agreement between the Colombian government and the Revolutionary Armed Forces of Colombia (FARC) was indeed reached in 2016. The peace deal aimed to end the decades-long armed conflict in Colombia. The agreement included a ceasefire and paved the way for negotiations on various issues, such as disarmament, political participation, and rural development. Therefore, the answer is: \textbf{B. Real}.\\
			\bottomrule
	\end{tabular}}
        \caption{Sample ``689529955294777344'' in the Twitter dataset. GPT 3.5 confirms the news is real.}
	\label{tab:exC02}
\end{table*}

\begin{table*}[!ht]
	\centering
	\resizebox{\linewidth}{!}
	{\begin{tabular}{l|p{16cm}}
			\toprule
Prompt &
There is a piece of news: ``\textcolor{blue}{RIP Kenny Baker, the actor behind legendary Star Wars droid R2-D2.}'' Verify the credibility of the news. Please choose the answer from the following options: A. Fake, B. Real.\\
			\midrule
Output &
As of my last knowledge update in January 2022, Kenny Baker, the actor who portrayed R2-D2 in the Star Wars films, passed away in 2016. Therefore, the answer is: \textbf{B. Real}.\\
			\bottomrule
	\end{tabular}}
        \caption{Sample ``764505291853627392'' in the Twitter dataset. GPT 3.5 confirms the news is real.}
	\label{tab:exC03}
\end{table*}

\begin{table*}[!ht]
	\centering
	\resizebox{\linewidth}{!}
	{\begin{tabular}{l|p{16cm}}
			\toprule
Prompt of Step 1 &
There is a piece of news: ``\textcolor{blue}{The State Council will officially approve Shanghai's request to abolish the establishment of Jing'an District and divide the original administrative regions of Jing'an District into Changning District and Huangpu District in the near future.}'' There are comments for the news: ``\textcolor{blue}{Changning District has expanded, and we hope for the great development of Changning culture!!!}'' ``\textcolor{blue}{But I really like Jing'an District, the most beautiful district name, don't merge it!}'' ``\textcolor{blue}{Really?}'' ``\textcolor{blue}{I don't understand, but it seems reasonable.}'' ``\textcolor{blue}{No way, right?!}'' ``\textcolor{blue}{Is Jing'an District going to change its past tense???}'' ``\textcolor{blue}{Didn't we already refute the rumors a few months ago}'' …… You need to do: (1) Based on the writing style and the commonsense knowledge, estimate the credibility of the news. (2) Based on the comments, analyze whether there are any rebuttals or conflicts, and then accordingly verify the authenticity of the news. Based on above results, please choose the answer from the following options: A. Fake, B. Real.\\
			\midrule
Output of Step 1 &
The news \textcolor{red}{\underline{seems to be presented as a statement of fact}}, mentioning that the State Council will officially approve Shanghai's request to abolish the establishment of Jing'an District. However, it lacks specific details or sources. The use of phrases like ``in the near future'' might suggest \textcolor{red}{\underline{a lack of concrete information or immediacy}}. The comments reflect a variety of opinions, ranging from support for the development of Changning District to concerns about losing the name ``Jing'an District.'' Some comments express \textcolor{red}{\underline{confusion or skepticism about the news}}, indicating a lack of clarity or consensus among the readers. The comment mentioning \textcolor{red}{\underline{the refutation of rumors a few months ago raises questions}} \textcolor{red}{\underline{about the consistency of the information}}. Therefore, the answer is: \textbf{A. Fake}.\\
\midrule
Prompt of Step 2 &
There is a piece of news: ``\textcolor{blue}{The State Council will officially approve Shanghai's request to abolish the establishment of Jing'an District and divide the original administrative regions of Jing'an District into Changning District and Huangpu District in the near future.}'' There are comments for the news: ``\textcolor{blue}{No~I object}'' ``\textcolor{blue}{Why merge?}'' ``\textcolor{blue}{At that time, the government even stepped forward to refute rumors}'' ``\textcolor{blue}{I saw this post three months ago.}'' ``\textcolor{blue}{I have become a resident of Changning District}'' ``\textcolor{blue}{What is the significance of it}'' ``\textcolor{blue}{Such fake news would make so many people believe it!}'' …… You need to do: (1) Based on the writing style and the commonsense knowledge, estimate the credibility of the news. (2) Based on the comments, analyze whether there are any rebuttals or conflicts, and then accordingly verify the authenticity of the news. Based on above results, please choose the answer from the following options: A. Fake, B. Real.\\
			\midrule
Output of Step 2 &
The news is presented as a statement, but it \textcolor{red}{\underline{lacks specific details or sources}}, making it challenging to verify its accuracy. Phrases like ``in the near futur'' suggest \textcolor{red}{\underline{a lack of immediacy and concrete information}}. The comments include objections, questions about the merger, references to previous government refutations, and skepticism about the news, indicating potential credibility issues. Several comments express objections, \textcolor{red}{\underline{questioning the reasons}} behind the merger and \textcolor{red}{\underline{citing past government refutations}}. A comment mentions having seen a similar post three months ago, suggesting \textcolor{red}{\underline{a lack of novelty or consistency in the information}}. One commenter claims to have become a resident of Changning District, potentially providing a firsthand perspective. Therefore, the answer is: \textbf{A. Fake}.\\
			\midrule
Prompt of Step 3 &
There is a piece of news: ``\textcolor{blue}{The State Council will officially approve Shanghai's request to abolish the establishment of Jing'an District and divide the original administrative regions of Jing'an District into Changning District and Huangpu District in the near future.}'' There are comments for the news: ``\textcolor{blue}{Jing'an District is going to become history}'' ``\textcolor{blue}{The trend is inevitable, we'll just wait for time.}'' ``\textcolor{blue}{Well, I've heard of this specific plan, but the problem is it's already a month now, how can we make it.}'' ``\textcolor{blue}{Really?}'' ``\textcolor{blue}{Fake, at least currently fake.}'' ``\textcolor{blue}{This is not true, is it...}'' ``\textcolor{blue}{Nonsense.}'' …… You need to do: (1) Based on the writing style and the commonsense knowledge, estimate the credibility of the news. (2) Based on the comments, analyze whether there are any rebuttals or conflicts, and then accordingly verify the authenticity of the news. Based on above results, please choose the answer from the following options: A. Fake, B. Real.\\
			\midrule
Output of Step 3 &
The news is presented as a statement, asserting that the State Council will officially approve Shanghai’s request to abolish Jing’an District and divide it into Changning District and Huangpu District. Phrases like ``Jing’an District is going to become history'' and ``The trend is inevitable'' suggest \textcolor{red}{\underline{a certain level of acceptance or resignation}}. Comments expressing doubts like ``Fake, at least currently fake'' and ``This is not true, is it...'' \textcolor{red}{\underline{indicate skepticism}}. The comment labeling it as ``Fake, at least currently fake'' and others \textcolor{red}{\underline{expressing disbelief}} or calling it ``Nonsense'' add weight to doubts about the authenticity. Therefore, the answer is: \textbf{A. Fake}.\\
			\midrule
Final Output &
The final estimation of the veracity of the news is: \textbf{The news is fake.}\\
			\bottomrule
	\end{tabular}}
        \caption{Sample ``3498854609883720'' in the Weibo dataset. LeRuD correctly predicts it as fake news.}
	\label{tab:EX-3498854609883720}
\end{table*}

\begin{table*}[!ht]
	\centering
	\resizebox{\linewidth}{!}
	{\begin{tabular}{l|p{16cm}}
			\toprule
Prompt of Step 1 &
There is a piece of news: ``\textcolor{blue}{Famous actor Morgan Freeman passed away in Los Angeles at 8am local time on Tuesday, February 19th due to heart disease.}'' There are comments for the news: ``\textcolor{blue}{A great actor!}'' ``\textcolor{blue}{Blessing.}'' ``\textcolor{blue}{So suddenly.}'' ``\textcolor{blue}{One of my favorite actors...}'' ``\textcolor{blue}{Oh my goodness, I really like him!}'' ``\textcolor{blue}{A great actor}'' ``\textcolor{blue}{I hope it's fake news!}'' …… You need to do: (1) Based on the writing style and the commonsense knowledge, estimate the credibility of the news. (2) Based on the comments, analyze whether there are any rebuttals or conflicts, and then accordingly verify the authenticity of the news. Based on above results, please choose the answer from the following options: A. Fake, B. Real.\\
			\midrule
Output of Step 1 &
The news about Morgan Freeman's death seems credible based on the provided information. The details include \textcolor{red}{\underline{the specific time, cause of death, and location}}, which adds \textcolor{red}{\underline{a level of specificity often found}} \textcolor{red}{\underline{in legitimate news reports}}. Additionally, the news is \textcolor{red}{\underline{written in a straightforward and factual manner}}, contributing to its credibility. The comments express sentiments of admiration and surprise, \textcolor{red}{\underline{with no apparent}} \textcolor{red}{\underline{rebuttals or conflicts}}. However, the comment "I hope it’s fake news!" \textcolor{red}{\underline{suggests some skepticism}}. It's important to note that public figures' deaths can sometimes be subject to rumors or hoaxes, but without additional information, it's challenging to determine the veracity solely based on the comments. Therefore, the answer is: \textbf{B. Real}.\\
\midrule
Prompt of Step 2 &
There is a piece of news: ``\textcolor{blue}{Famous actor Morgan Freeman passed away in Los Angeles at 8am local time on Tuesday, February 19th due to heart disease.}'' There are comments for the news: ``\textcolor{blue}{Sincerely hope it's spreading rumors}'' ``\textcolor{blue}{Really? The list of last wishes is also funny.}'' ``\textcolor{blue}{I was deceived! Rumors!}'' ``\textcolor{blue}{I can't believe it!}'' ``\textcolor{blue}{I will continue to watch your movies.}'' ``\textcolor{blue}{Is it fake news? Has someone refuted the rumor?}'' ``\textcolor{blue}{Morgan Freeman is said to be died again?}'' …… You need to do: (1) Based on the writing style and the commonsense knowledge, estimate the credibility of the news. (2) Based on the comments, analyze whether there are any rebuttals or conflicts, and then accordingly verify the authenticity of the news. Based on above results, please choose the answer from the following options: A. Fake, B. Real.\\
			\midrule
Output of Step 2 &
The news mentions \textcolor{red}{\underline{specific details like the time, cause of death, and location}}, which lends an air of credibility. However, there is a mix of reactions in the comments, with some \textcolor{red}{\underline{expressing doubt and}} \textcolor{red}{\underline{questioning the authenticity}} of the news. Some comments express skepticism or hope that it's a rumor, indicating potential disbelief in the news. The comment ``Morgan Freeman is said to be died again?'' \textcolor{red}{\underline{directly questions the veracity of the news, indicating a potential conflict or doubt}}. Therefore, the answer is: \textbf{A. Fake}.\\
			\midrule
Prompt of Step 3 &
There is a piece of news: ``\textcolor{blue}{Famous actor Morgan Freeman passed away in Los Angeles at 8am local time on Tuesday, February 19th due to heart disease.}'' There are comments for the news: ``\textcolor{blue}{Fake news!}'' ``\textcolor{blue}{I'sorry to hear that...}'' ``\textcolor{blue}{Why do so many people like to say that others have died~~~}'' ``\textcolor{blue}{Just reposted it and realized it's fake news!!!}'' ``\textcolor{blue}{It has been identified as fake news, please don't believe it}'' ``\textcolor{blue}{RIP.}'' ``\textcolor{blue}{This has been confirmed as a rumor.}'' …… You need to do: (1) Based on the writing style and the commonsense knowledge, estimate the credibility of the news. (2) Based on the comments, analyze whether there are any rebuttals or conflicts, and then accordingly verify the authenticity of the news. Based on above results, please choose the answer from the following options: A. Fake, B. Real.\\
			\midrule
Output of Step 3 &
The presence of comments like "Fake news!" and "It has been identified as fake news, please don’t believe it" \textcolor{red}{\underline{strongly suggests a lack of credibility}}. The comment expressing realization that it's fake after reposting also adds to the skepticism surrounding the news. The majority of comments \textcolor{red}{\underline{explicitly label the news as fake and provide reasons for their skepticism}}. Phrases like "confirmed as a rumor" and "identified as fake news" indicate \textcolor{red}{\underline{clear rebuttals and conflicts}} with the authenticity of the news. There is an absence of supportive comments or credible sources affirming the news. Therefore, the answer is: \textbf{A. Fake}.\\
			\midrule
Final Output &
The final estimation of the veracity of the news is: \textbf{The news is fake.}\\
			\bottomrule
	\end{tabular}}
        \caption{Sample ``3547977337564354'' in the Weibo dataset. LeRuD correctly predicts it as fake news.}
	\label{tab:EX-3547977337564354}
\end{table*}

\begin{table*}[!ht]
	\centering
	\resizebox{\linewidth}{!}
	{\begin{tabular}{l|p{16cm}}
			\toprule
Prompt of Step 1 &
There is a piece of news: ``\textcolor{blue}{Today, Shanghai Metro Line 10 experienced a malfunction, and the two trains successfully merged, which is truly terrifying.}'' There are comments for the news: ``\textcolor{blue}{What's wrong with Line 10 again,....}'' ``\textcolor{blue}{When? I also took Line 10 today}'' ``\textcolor{blue}{What? How was this achieved?}'' ``\textcolor{blue}{Really?}'' ``\textcolor{blue}{It's not surprising anymore}'' ``\textcolor{blue}{This is really funny.}'' ``\textcolor{blue}{I think it's the same as the last picture. Is it true?}'' …… You need to do: (1) Based on the writing style and the commonsense knowledge, estimate the credibility of the news. (2) Based on the comments, analyze whether there are any rebuttals or conflicts, and then accordingly verify the authenticity of the news. Based on above results, please choose the answer from the following options: A. Fake, B. Real.\\
			\midrule
Output of Step 1 &
The news itself \textcolor{red}{\underline{seems unusual}}, as it claims that two trains on Shanghai Metro Line 10 successfully merged due to a malfunction. This scenario \textcolor{red}{\underline{raises questions about the technical feasibility}} and safety of such an event. The comments express \textcolor{red}{\underline{a mix of skepticism, concern, and curiosity}}. Some users question the reliability of Line 10, while others seek more information about the incident. The range of reactions suggests a level of uncertainty among the audience. Without additional details or official sources confirming the incident, it's \textcolor{red}{\underline{difficult to verify the authenticity of the news}}. Therefore, the answer is: \textbf{A. Fake}.\\
\midrule
Prompt of Step 2 &
There is a piece of news: ``\textcolor{blue}{Today, Shanghai Metro Line 10 experienced a malfunction, and the two trains successfully merged, which is truly terrifying.}'' There are comments for the news: ``\textcolor{blue}{Don't use the word terror indiscriminately, it's just that the train ahead is malfunctioning. The train behind will push forward for rescue, connect the malfunctioning train, and push it to the designated location to restore the main line operation as soon as possible! This is just a commonly used train rescue method in the subway, there is no such thing as horror.}'' ``\textcolor{blue}{Terrifying... Marketing Weibo attracts attention from the public}'' ``\textcolor{blue}{I haven't seen it either...}'' ``\textcolor{blue}{Scared me...}'' ``\textcolor{blue}{No knowledge, it's really scary!}'' ``\textcolor{blue}{Whoever spreads rumors is terrifying!}'' ``\textcolor{blue}{Iron Brave Train Warrior, transform!}'' …… You need to do: (1) Based on the writing style and the commonsense knowledge, estimate the credibility of the news. (2) Based on the comments, analyze whether there are any rebuttals or conflicts, and then accordingly verify the authenticity of the news. Based on above results, please choose the answer from the following options: A. Fake, B. Real.\\
			\midrule
Output of Step 2 &
The comment that explains the situation mentions that the merging of trains is a commonly used rescue method in the subway when there is a malfunction. \textcolor{red}{\underline{This comment provides a plausible explanation for}} \textcolor{red}{\underline{the incident, suggesting that it might not be as terrifying as the initial news headline implies.}} Other comments \textcolor{red}{\underline{express skepticism or confusion}}, with some users \textcolor{red}{\underline{suggesting that the use of the word ``terrifying''}} \textcolor{red}{\underline{might be for marketing purposes or to attract public attention}}. The explanation provided in one of the comments aligns with common subway rescue procedures, making the incident more plausible and less alarming. The skepticism expressed in some comments, coupled with the suggestion of marketing tactics, \textcolor{red}{\underline{raises questions about the accuracy and sensationalism}} of the initial news report. Therefore, the answer is: \textbf{A. Fake}.\\
			\midrule
Prompt of Step 3 &
There is a piece of news: ``\textcolor{blue}{Today, Shanghai Metro Line 10 experienced a malfunction, and the two trains successfully merged, which is truly terrifying.}'' There are comments for the news: ``\textcolor{blue}{It's too scary. It seems like we can't sit on Line 10.}'' ``\textcolor{blue}{Recently, the subway has been broken down frequently.}'' ``\textcolor{blue}{I personally believe that facing this kind of marketing and Weibo popularization, which takes spreading rumors as its own responsibility, is futile no matter how much it is.}'' ``\textcolor{blue}{Please allow me to scold you. Such a simple matter is like a car malfunctioning and being towed back for repair. I'm amazed at your understanding.}'' ``\textcolor{blue}{This is a very normal rescue method, don't use the word terror in the title party.}'' ``\textcolor{blue}{Really?}'' ``\textcolor{blue}{Rescue! make a big fuss!}'' …… You need to do: (1) Based on the writing style and the commonsense knowledge, estimate the credibility of the news. (2) Based on the comments, analyze whether there are any rebuttals or conflicts, and then accordingly verify the authenticity of the news. Based on above results, please choose the answer from the following options: A. Fake, B. Real.\\
			\midrule
Output of Step 3 &
Comments express \textcolor{red}{\underline{a mix of concern, skepticism, and criticism}} toward the news. Some users find the news terrifying and express reluctance to use Line 10, while others criticize the use of the word ``terror'' and suggest that it's a normal rescue method. A few comments \textcolor{red}{\underline{provide a more rational perspective,}} \textcolor{red}{\underline{likening the situation to a car malfunction being towed for repair, and expressing amazement at the mis-}} \textcolor{red}{\underline{understanding}}. The comments present a diversity of opinions, with some users expressing genuine concern and others debunking the severity of the situation. \textcolor{red}{\underline{Rational explanations in the comments suggest that}} \textcolor{red}{\underline{the incident might be a routine and safe subway rescue operation, contrary to the alarming tone of the}} \textcolor{red}{\underline{news.}} Therefore, the answer is: \textbf{A. Fake}.\\
			\midrule
Final Output &
The final estimation of the veracity of the news is: \textbf{The news is fake.}\\
			\bottomrule
	\end{tabular}}
        \caption{Sample ``3465261540071546'' in the Weibo dataset. LeRuD correctly predicts it as fake news.}
	\label{tab:EX-3465261540071546}
\end{table*}

\begin{table*}[!ht]
	\centering
	\resizebox{\linewidth}{!}
	{\begin{tabular}{l|p{16cm}}
			\toprule
Prompt of Step 1 &
There is a piece of news: ``\textcolor{blue}{NHL expansion – four teams added, Quebec City, Toronto, Seattle, and Las Vegas \$1.4b in expansion fees}'' There are comments for the news: ``\textcolor{blue}{Short term thinking...  uggh}'' ``\textcolor{blue}{Phoenix is not moving and sadly neither is Miami}'' ``\textcolor{blue}{Hopefully this also helps Seattle get an NBA team back.}'' ``\textcolor{blue}{Is this done deal? Or is this just speculation? Thank you}'' ``\textcolor{blue}{Has 3-4 yrs to build up talent. Can be handled if KHL stops taking guys.}'' ``\textcolor{blue}{Seattle has one Stanley Cup already. Woot! Lets get another.}'' ``\textcolor{blue}{YES!!! They need a facility though!}'' …… You need to do: (1) Based on the writing style and the commonsense knowledge, estimate the credibility of the news. (2) Based on the comments, analyze whether there are any rebuttals or conflicts, and then accordingly verify the authenticity of the news. Based on above results, please choose the answer from the following options: A. Fake, B. Real.\\
			\midrule
Output of Step 1 &
The \textcolor{red}{\underline{writing style of the news seems plausible}}, mentioning \textcolor{red}{\underline{specific details such as the addition of four}} \textcolor{red}{\underline{NHL teams}} (Quebec City, Toronto, Seattle, and Las Vegas) and the associated \$1.4 billion in expansion fees. The information appears coherent and consistent, suggesting a level of credibility. However, without additional sources or context, it's challenging to make a definitive judgment. The comments reflect a mix of opinions and concerns. Some express excitement about the expansion, while others express skepticism or raise questions about specific locations like Phoenix and Miami. The comment ``Has 3-4 yrs to build up talent. Can be handled if KHL stops taking guys''implies \textcolor{red}{\underline{considerations about player}} \textcolor{red}{\underline{talent development and potential challenges related to the KHL (Kontinental Hockey League)}}. There is \textcolor{red}{\underline{no direct rebuttal or conflict}} in the comments that would immediately discredit the news. Therefore, the answer is: \textbf{B. Real}.\\
\midrule
Prompt of Step 2 &
There is a piece of news: ``\textcolor{blue}{NHL expansion – four teams added, Quebec City, Toronto, Seattle, and Las Vegas \$1.4b in expansion fees}'' There are comments for the news: ``\textcolor{blue}{you think the Seattle City Council will go with an NHL 1st option for the Sodo Arena? I doubt it but I hope so}'' ``\textcolor{blue}{I don't take him seriously.  Also implies that Seattle can't have NHL AND NBA... gotta be one or the other.}'' ``\textcolor{blue}{if this is true then wouls arena be built prior to sea getting a team?}'' ``\textcolor{blue}{Bettman said himself a few weeks ago that Seattle not being considered right now. If they build an arena? Sure. But not before that.}'' ``\textcolor{blue}{Seriously?}'' ``\textcolor{blue}{Toronto has a team.....}'' ``\textcolor{blue}{there will be an imbalance in schedule and conferences/divisions. There's no way.}'' …… You need to do: (1) Based on the writing style and the commonsense knowledge, estimate the credibility of the news. (2) Based on the comments, analyze whether there are any rebuttals or conflicts, and then accordingly verify the authenticity of the news. Based on above results, please choose the answer from the following options: A. Fake, B. Real.\\
			\midrule
Output of Step 2 &
The news mentions the expansion of four NHL teams (Quebec City, Toronto, Seattle, and Las Vegas), along with the associated \$1.4 billion in expansion fees. The comments \textcolor{red}{\underline{express skepticism and doubts}} about the news. One comment \textcolor{red}{\underline{questions whether the Seattle City Council}} \textcolor{red}{\underline{would prioritize an NHL option for the Sodo Arena}}, and another doubts the credibility of the source. The comment ``Toronto has a team.....'' \textcolor{red}{\underline{points out a factual error in the news, as Toronto already had an NHL}} \textcolor{red}{\underline{team before the mentioned expansion}}. These comments indicate a level of suspicion and raise valid concerns about the accuracy of the news. Therefore, the answer is: \textbf{A. Fake}.\\
			\midrule
Prompt of Step 3 &
There is a piece of news: ``\textcolor{blue}{NHL expansion – four teams added, Quebec City, Toronto, Seattle, and Las Vegas \$1.4b in expansion fees}'' There are comments for the news: ``\textcolor{blue}{I understand. But the Maple Leafs have been there for ages. Who will not stay on as a fan for them?}'' ``\textcolor{blue}{Seriously, i think we can skip Toronto. One is enough}'' ``\textcolor{blue}{I don't know why they'd send another team to Toronto. The broke the first one. Awful organization.}'' ``\textcolor{blue}{Doubt it}'' ``\textcolor{blue}{why? we don't have the market size.}'' ``\textcolor{blue}{source? Hallucination?}'' ``\textcolor{blue}{You're saying this and the \#NHL is saying no.  Which one of you is lying?}'' …… You need to do: (1) Based on the writing style and the commonsense knowledge, estimate the credibility of the news. (2) Based on the comments, analyze whether there are any rebuttals or conflicts, and then accordingly verify the authenticity of the news. Based on above results, please choose the answer from the following options: A. Fake, B. Real.\\
			\midrule
Output of Step 3 &
The news mentions the expansion of four NHL teams (Quebec City, Toronto, Seattle, and Las Vegas) with \$1.4 billion in expansion fees. The comments \textcolor{red}{\underline{express skepticism and doubts}} about the news. Multiple comments \textcolor{red}{\underline{question the inclusion of Toronto in the expansion, citing the Maple Leafs' long-standing}} \textcolor{red}{\underline{presence in the NHL}}. There are also\textcolor{red}{\underline{ negative opinions about sending another team to Toronto}}, with comments criticizing the organization or \textcolor{red}{\underline{expressing doubts about the market size}}. The comment ``You’re saying this and the NHL is saying no. Which one of you is lying?'' \textcolor{red}{\underline{directly challenges the credibility}} of the news source. Therefore, the answer is: \textbf{A. Fake}.\\
			\midrule
Final Output &
The final estimation of the veracity of the news is: \textbf{The news is fake.}\\
			\bottomrule
	\end{tabular}}
        \caption{Sample ``504448356584677377'' in the Twitter dataset. LeRuD correctly predicts it as fake news.}
	\label{tab:EX-504448356584677377}
\end{table*}

\begin{table*}[!ht]
	\centering
	\resizebox{\linewidth}{!}
	{\begin{tabular}{l|p{16cm}}
			\toprule
Prompt of Step 1 &
There is a piece of news: ``\textcolor{blue}{Months ago, Rupert Murdoch bought National Geographic and fired most of writers. This just hit the stands.}'' There are comments for the news: ``\textcolor{blue}{Hmm. Doesn't look like the kind of cover one wd expect from the NG. And I don't mean that in a good way.}'' ``\textcolor{blue}{is there anything he doesn't ruin?!}'' ``\textcolor{blue}{this is something sad to learn of. NatGeo was one of the last.}'' ``\textcolor{blue}{This isn't true, it's an old cover that's doing the rounds}'' ``\textcolor{blue}{noooooooo! That's so dumb.}'' ``\textcolor{blue}{I cancelled my NG subscription years ago. The writing then wasn't much better.}'' ``\textcolor{blue}{Murdock keeping his stupid sheep in line with ANYTHING but Truth! If there is a hell he should rot there!}'' …… You need to do: (1) Based on the writing style and the commonsense knowledge, estimate the credibility of the news. (2) Based on the comments, analyze whether there are any rebuttals or conflicts, and then accordingly verify the authenticity of the news. Based on above results, please choose the answer from the following options: A. Fake, B. Real.\\
			\midrule
Output of Step 1 &
Based on the writing style and common sense knowledge, it's important to note that the news lacks specific details such as the source, date, and specific actions taken by Rupert Murdoch. The language used in the news is \textcolor{red}{\underline{somewhat informal and lacks the journalistic tone}} often found in credible news sources. Additionally, the \textcolor{red}{\underline{absence of concrete information}} about the event raises concerns about its credibility. Analyzing the comments, there is a mix of reactions. Some express skepticism, questioning the authenticity of the news, while others react emotionally with disappointment and anger. One comment even suggests that the news might be based on an old cover that is circulating. The varied reactions indicate \textcolor{red}{\underline{a lack of consensus among readers}}. Therefore, the answer is: \textbf{A. Fake}.\\
			\midrule
Final Output &
The final estimation of the veracity of the news is: \textbf{The news is fake.}\\
			\bottomrule
	\end{tabular}}
        \caption{Sample ``676120162018451456'' in the Twitter dataset. LeRuD correctly predicts it as fake news.}
	\label{tab:EX-676120162018451456}
\end{table*}

\begin{table*}[!ht]
	\centering
	\resizebox{\linewidth}{!}
	{\begin{tabular}{l|p{16cm}}
			\toprule
Prompt of Step 1 &
There is a piece of news: ``\textcolor{blue}{Mike Brown murder trial witness, \#DeAndreJoshua fatally shot and set on fire after verdict.}'' There are comments for the news: ``\textcolor{blue}{This is ridiculous. It's time we take a stand}'' ``\textcolor{blue}{Your article doesn't say anything about him being a witness. Can you link to that fact?}'' ``\textcolor{blue}{The article doesn't mention he was a witness, got a source? Let the speculation begin.}'' ``\textcolor{blue}{Speechless.}'' ``\textcolor{blue}{Google his name and read some articles. Few witnesses gave testimony for fear of this happening.}'' ``\textcolor{blue}{ I had initially done so, however was unable to locate anything substantial.. Only speculation.}'' ``\textcolor{blue}{I'm sure the witnesses were told to keep quiet to avoid situations like this. Not sure if any will be confirmed}'' …… You need to do: (1) Based on the writing style and the commonsense knowledge, estimate the credibility of the news. (2) Based on the comments, analyze whether there are any rebuttals or conflicts, and then accordingly verify the authenticity of the news. Based on above results, please choose the answer from the following options: A. Fake, B. Real.\\
			\midrule
Output of Step 1 &
The news report mentions a shocking incident involving the murder trial witness, DeAndre Joshua, being fatally shot and set on fire after the verdict. The writing style \textcolor{red}{\underline{doesn't provide specific details about the}} \textcolor{red}{\underline{incident or cite any credible sources}}. The information \textcolor{red}{\underline{lacks concrete evidence and seems sensational-}} \textcolor{red}{\underline{ized}}, which raises concerns about its credibility. The comments \textcolor{red}{\underline{highlight skepticism and a lack of con-}} \textcolor{red}{\underline{crete evidence or sources}}. Some comments express skepticism and request sources to confirm the claim that DeAndre Joshua was a witness. One comment mentions that few witnesses gave testimony for fear of such incidents, indicating \textcolor{red}{\underline{a potential environment of intimidation}}. Therefore, the answer is: \textbf{A. Fake}.\\
			\midrule
Final Output &
The final estimation of the veracity of the news is: \textbf{The news is fake.}\\
			\bottomrule
	\end{tabular}}
        \caption{Sample ``538412662858588160'' in the Twitter dataset. LeRuD correctly predicts it as fake news.}
	\label{tab:EX-538412662858588160}
\end{table*}

\begin{table*}[!ht]
	\centering
	\resizebox{\linewidth}{!}
	{\begin{tabular}{l|p{16cm}}
			\toprule
Prompt of Step 1 &
There is a piece of news: ``\textcolor{blue}{Jay Z and Beyoncé want to buy rights to the Confederate flag}'' There are comments for the news: ``\textcolor{blue}{this isn't true. lol. do better, Complex.}'' ``\textcolor{blue}{huh???}'' ``\textcolor{blue}{lies!!}'' ``\textcolor{blue}{wut? RT Jay Z and Beyoncé want to buy rights to the Confederate}'' ``\textcolor{blue}{When you got it RT Jay and Beyoncé want to buy rights to the Confederate flag}'' ``\textcolor{blue}{idek wut dis mean RT Jay Z and Beyoncé want to buy rights to the Confederate flag}'' ``\textcolor{blue}{Whatever man. The stock exchange just halted trading...screw that flag}'' …… You need to do: (1) Based on the writing style and the commonsense knowledge, estimate the credibility of the news. (2) Based on the comments, analyze whether there are any rebuttals or conflicts, and then accordingly verify the authenticity of the news. Based on above results, please choose the answer from the following options: A. Fake, B. Real.\\
			\midrule
Output of Step 1 &
The writing style of the comments \textcolor{red}{\underline{suggests skepticism and confusion regarding the news}}. Phrases like ``lol,'' ``huh???'' and ``wut?'' indicate \textcolor{red}{\underline{a lack of belief or understanding of the information}}. Additionally, some comments \textcolor{red}{\underline{express disbelief or question the authenticity}} of the news, such as ``lies!!'' and ``do better, Complex.'' Overall, the tone of the comments \textcolor{red}{\underline{leans towards skepticism}}. The comments contain a mix of reactions, with some \textcolor{red}{\underline{expressing doubt or disbelief}}, while others seem to accept or support the news. It's essential to note that comments alone may not provide a definitive answer to the authenticity of the news. However, the skepticism expressed in some comments suggests that there might be conflicts or differing opinions among the audience. Based on the above analysis, the news appears to be more likely fake or at least questionable. The comments \textcolor{red}{\underline{reflect a degree of doubt and confusion}}, pointing towards a lack of credibility. Therefore, the answer is: \textbf{A. Fake}.\\
			\midrule
Final Output &
The final estimation of the veracity of the news is: \textbf{The news is fake.}\\
			\bottomrule
	\end{tabular}}
        \caption{Sample ``618805892222468096'' in the Twitter dataset. LeRuD correctly predicts it as fake news.}
	\label{tab:EX-618805892222468096}
\end{table*}

\begin{table*}[!ht]
	\centering
	\resizebox{\linewidth}{!}
	{\begin{tabular}{l|p{16cm}}
			\toprule
Prompt of Step 1 &
There is a piece of news: ``\textcolor{blue}{The Panjin Public Security Bureau is investigating itself again. Villager Wang Shujie was shot and killed by police officer Zhang Yan, but the investigation found that the police shot legally! Li Hongna was gang raped by the director of Panjin Police Station and local ruffians. Faced with 30000 reposts from netizens, the Panjin Public Security Bureau couldn't sit still and responded to netizens by posting Weibo messages online, urging everyone to be patient and wait for the investigation results. After so many days, the girl's two distress messages were deleted. Where did you lock her up?}'' There are comments for the news: ``\textcolor{blue}{Repost}'' ``\textcolor{blue}{Seeking the truth.}'' ``\textcolor{blue}{No wonder I always feel so familiar when I hear the place name Panjin while watching a movie today!}'' ``\textcolor{blue}{It is only natural that Wang Shujie was killed. He first illegally injured the construction workers, then attacked the police with a knife and attempted to attack even after the police fired three shots into the sky. Li Hongna's case has been investigated for a long time, and it was a false accusation. All the lies she said were unfounded. Li is a drug addict and has been falsely accused by the police on Weibo.}'' ``\textcolor{blue}{angry}'' ``\textcolor{blue}{angry}'' ``\textcolor{blue}{Panjin Public Security is in chaos.}'' …… You need to do: (1) Based on the writing style and the commonsense knowledge, estimate the credibility of the news. (2) Based on the comments, analyze whether there are any rebuttals or conflicts, and then accordingly verify the authenticity of the news. Based on above results, please choose the answer from the following options: A. Fake, B. Real.\\
			\midrule
Output of Step 1 &
The news contains \textcolor{red}{\underline{conflicting information and seems to be presented in a sensationalized manner}}. The \textcolor{red}{\underline{tone of the news raises suspicions}}, and the use of phrases like ``investigating itself again'' and ``couldn't sit still'' \textcolor{red}{\underline{may indicate bias}}. Additionally, the comments show a range of opinions, including both support and skepticism. Without further evidence or reliable sources, it is challenging to determine the credibility of the news. However, the news seems to \textcolor{red}{\underline{lack specific details and relies heavily on emotional language}}, which could be indicative of a less credible source. The comments reflect a mix of opinions, with some expressing skepticism and others defending the actions of the Panjin Public Security Bureau. The \textcolor{red}{\underline{conflicting narratives in the comments suggest a lack of consensus among the readers}}. Some comments \textcolor{red}{\underline{question the legitimacy of the claims}}, while others show support for the authorities. There is no clear consensus among the commenters. Therefore, the answer is: \textbf{A. Fake}.\\
			\midrule
Final Output &
The final estimation of the veracity of the news is: \textbf{The news is fake.}\\
			\bottomrule
	\end{tabular}}
        \caption{Sample ``3507985437364629'' in the Weibo dataset. LeRuD correctly predicts it as fake news.}
	\label{tab:EX-3507985437364629}
\end{table*}

\begin{table*}[!ht]
	\centering
	\resizebox{\linewidth}{!}
	{\begin{tabular}{l|p{16cm}}
			\toprule
Prompt of Step 1 &
There is a piece of news: ``\textcolor{blue}{The Paris attack has sounded a global anti-terrorism alert.}'' There are comments for the news: ``\textcolor{blue}{Shocked, just heard!}'' ``\textcolor{blue}{The cause of terrorist attacks is not due to the general public, but they bear the tragic consequences}'' ``\textcolor{blue}{Only wish for world peace!}'' ``\textcolor{blue}{Terrorism will always be the common enemy of all humanity, and any violent attacks against civilians should be condemned and countered. Mourning for Paris.}'' ``\textcolor{blue}{The road to counter-terrorism is still long, and there is still much we need to do.}'' ``\textcolor{blue}{Rest in peace for the deceased.}'' ``\textcolor{blue}{Civilian people are innocent.}'' …… You need to do: (1) Based on the writing style and the commonsense knowledge, estimate the credibility of the news. (2) Based on the comments, analyze whether there are any rebuttals or conflicts, and then accordingly verify the authenticity of the news. Based on above results, please choose the answer from the following options: A. Fake, B. Real.\\
			\midrule
Output of Step 1 &
Based on the information provided, it's challenging to directly estimate the credibility of the news. The statement ``The Paris attack has sounded a global anti-terrorism alert'' is \textcolor{red}{\underline{quite general and lacks specific}} \textcolor{red}{\underline{details about the incident}}. Credibility assessment would require more information, such as the news source, additional context, and corroboration from reliable sources. Analyzing the comments, there is \textcolor{red}{\underline{no evidence of rebuttals or conflicts regarding the authenticity}} of the news. The comments express shock, sympathy, condemnation of terrorism, and a desire for world peace. The comments are \textcolor{red}{\underline{more focused on}} \textcolor{red}{\underline{expressing emotions and opinions rather than disputing the news itself}}. So the answer is undetermined.\\
\midrule
Prompt of Step 2 &
There is a piece of news: ``\textcolor{blue}{The Paris attack has sounded a global anti-terrorism alert.}'' There are comments for the news: ``\textcolor{blue}{Watching the report and listening to gunshots, my throat and eyes tightened, my eyes became sore, and everyone's eyes were filled with helplessness, filled with a sense of terror everywhere. Counter terrorism is urgent and cannot be ignored.}'' ``\textcolor{blue}{The world needs to work together to strike.}'' ``\textcolor{blue}{Let's pray.}'' ``\textcolor{blue}{Every country should attach importance to terrorist attacks and take them as a lesson.}'' ``\textcolor{blue}{Pray that this romantic city will no longer be harmed.}'' ``\textcolor{blue}{The Paris attack has sounded a global anti-terrorism alert.}'' ``\textcolor{blue}{May the deceased rest in peace! World peace!}'' …… You need to do: (1) Based on the writing style and the commonsense knowledge, estimate the credibility of the news. (2) Based on the comments, analyze whether there are any rebuttals or conflicts, and then accordingly verify the authenticity of the news. Based on above results, please choose the answer from the following options: A. Fake, B. Real.\\
			\midrule
Output of Step 2 &
Based on the provided information, the news seems to be credible and reflects a serious incident – a terrorist attack in Paris. The writing style is \textcolor{red}{\underline{coherent and consistent with typical news reports on such}} \textcolor{red}{\underline{events}}. The comments express shock, sadness, and a call for global unity against terrorism, which aligns with the expected public response to such incidents. There are \textcolor{red}{\underline{no apparent rebuttals or conflicts}} in the comments; rather, they \textcolor{red}{\underline{convey a shared sentiment of concern, grief for the victims, and a collective call}} \textcolor{red}{\underline{for action against terrorism}}. The comments do not raise doubts about the authenticity of the news but rather emphasize the need for global cooperation and solidarity. Therefore, the answer is: \textbf{B. Real}.\\
			\midrule
Prompt of Step 3 &
There is a piece of news: ``\textcolor{blue}{The Paris attack has sounded a global anti-terrorism alert.}'' There are comments for the news: ``\textcolor{blue}{May everyone in the world have a beautiful heart, and may there be no more disasters in the world.}'' ``\textcolor{blue}{Is it not good to get along well with others?}'' ``\textcolor{blue}{The casualties are continuing to expand, and the most heartbreaking thing is that countries are still debating countermeasures.}'' ``\textcolor{blue}{Blessing Paris, may the world be at peace.}'' ``\textcolor{blue}{I still don't want to go out. It feels unsafe everywhere abroad.}'' ``\textcolor{blue}{No matter what, innocent civilians should not be harmed.}'' ``\textcolor{blue}{Why are these terrorists so terrifying? May the world be peaceful, please let go of the kind and innocent people.}'' …… You need to do: (1) Based on the writing style and the commonsense knowledge, estimate the credibility of the news. (2) Based on the comments, analyze whether there are any rebuttals or conflicts, and then accordingly verify the authenticity of the news. Based on above results, please choose the answer from the following options: A. Fake, B. Real.\\
			\midrule
Output of Step 3 &
Based on the provided information, the news appears to be credible as it describes a serious event – a terrorist attack in Paris. The writing style is \textcolor{red}{\underline{consistent with typical news reporting on such incidents}}. Analyzing the comments, there are no evident rebuttals or conflicts. Instead, the comments express concerns, sympathy, and a desire for peace. They highlight the impact of the attack on innocent civilians and call for measures to ensure safety. These sentiments \textcolor{red}{\underline{align with the expected public response to a}} \textcolor{red}{\underline{tragic event like a terrorist attack}}. Therefore, the answer is: \textbf{B. Real}.\\
			\midrule
Final Output &
The final estimation of the veracity of the news is: \textbf{The news is real.}\\
			\bottomrule
	\end{tabular}}
        \caption{Sample ``3909084371985775'' in the Weibo dataset. LeRuD correctly predicts it as real news.}
	\label{tab:EX-3909084371985775}
\end{table*}

\begin{table*}[!ht]
	\centering
	\resizebox{\linewidth}{!}
	{\begin{tabular}{l|p{16cm}}
			\toprule
Prompt of Step 1 &
There is a piece of news: ``\textcolor{blue}{27 years later, she found a firefighter who sacrificed her life to save her.}'' There are comments for the news: ``\textcolor{blue}{They are all good examples.}'' ``\textcolor{blue}{Moved!}'' ``\textcolor{blue}{Give a thumbs up to the firefighting hero, and also give a thumbs up to the girl's gratitude!}'' ``\textcolor{blue}{A life-saving benefactor. Although it is his responsibility, I also want to express my gratitude.}'' ``\textcolor{blue}{Repost}'' ``\textcolor{blue}{Positive energy!}'' ``\textcolor{blue}{Diligence and understanding gratitude are both good virtues!}'' …… You need to do: (1) Based on the writing style and the commonsense knowledge, estimate the credibility of the news. (2) Based on the comments, analyze whether there are any rebuttals or conflicts, and then accordingly verify the authenticity of the news. Based on above results, please choose the answer from the following options: A. Fake, B. Real.\\
			\midrule
Output of Step 1 &
Based on the information provided, it's challenging to definitively determine the credibility of the news without additional context or specific details about the source. However, there are \textcolor{red}{\underline{some positive comments}} \textcolor{red}{\underline{expressing emotions like being moved, giving thumbs up, and acknowledging the heroism of the firefigh-}} \textcolor{red}{\underline{ter and the gratitude of the girl}}. These comments \textcolor{red}{\underline{generally support the authenticity}} of the news. Without any specific details or evidence to suggest otherwise, it's reasonable to lean towards a more positive assessment. Therefore, considering the writing style and the overall positive sentiment in the comments, the news appears to be more likely real than fake. Therefore, the answer is: \textbf{B. Real}.\\
\midrule
Prompt of Step 2 &
There is a piece of news: ``\textcolor{blue}{27 years later, she found a firefighter who sacrificed her life to save her.}'' There are comments for the news: ``\textcolor{blue}{Good people must have good karma!}'' ``\textcolor{blue}{With a grateful heart.}'' ``\textcolor{blue}{A good girl who knows how to be grateful, also has a good family education.}'' ``\textcolor{blue}{This spirit is worthy of encouragement, and a grateful heart will never be erased. The transmission and promotion of positive energy will not stop.}'' ``\textcolor{blue}{Being a firefighter for 27 years is not easy!}'' ``\textcolor{blue}{Sincere gratitude and gratitude in the heart are sometimes more precious than money!}'' ``\textcolor{blue}{A grateful person will have a good life.}'' …… You need to do: (1) Based on the writing style and the commonsense knowledge, estimate the credibility of the news. (2) Based on the comments, analyze whether there are any rebuttals or conflicts, and then accordingly verify the authenticity of the news. Based on above results, please choose the answer from the following options: A. Fake, B. Real.\\
			\midrule
Output of Step 2 &
Based on the information provided, the writing style and comments \textcolor{red}{\underline{generally seem positive and suppor-}} \textcolor{red}{\underline{tive of the news}}. The comments express sentiments of gratitude, encouragement, and recognition of the challenges faced by firefighters. There are \textcolor{red}{\underline{no apparent signs of rebuttals or conflicts}} in the comments. Given the positive nature of both the news and the comments, and without specific details suggesting otherwise, it appears more likely that the news is real. Therefore, the answer is: \textbf{B. Real}.\\
			\midrule
Prompt of Step 3 &
There is a piece of news: ``\textcolor{blue}{27 years later, she found a firefighter who sacrificed her life to save her.}'' There are comments for the news: ``\textcolor{blue}{A servant of the people, a good firefighter who does not seek anything in return; For 27 years, I will never forget to search for benefactors. My parents have taught me well, and I am grateful for the people and things around me!}'' ``\textcolor{blue}{So warm, Uncle Firefighter looks so kind.}'' ``\textcolor{blue}{Those who show gratitude are all good people.}'' ``\textcolor{blue}{The heroic product of sacrificing oneself to save babies, the true emotions in the world are deeply moving!}'' ``\textcolor{blue}{Sun Bingjing visited the fire brigade, and the retired squad leader inferred that he must have rich experience in rescuing people in a dangerous fire scene. Eventually, he found the 27 year old squad leader, Lu Yuming.}'' ``\textcolor{blue}{Never forget the kindness of saving lives.}'' ``\textcolor{blue}{A good person lives a peaceful life.}'' …… You need to do: (1) Based on the writing style and the commonsense knowledge, estimate the credibility of the news. (2) Based on the comments, analyze whether there are any rebuttals or conflicts, and then accordingly verify the authenticity of the news. Based on above results, please choose the answer from the following options: A. Fake, B. Real.\\
			\midrule
Output of Step 3 &
Based on the information provided, the writing style and comments continue to convey positive sentiments, expressing gratitude, admiration for the firefighter, and recognition of the importance of showing appreciation. \textcolor{red}{\underline{There is a specific mention of Sun Bingjing visiting the fire brigade and finding the 27-year-old}} \textcolor{red}{\underline{squad leader, Lu Yuming, which adds detail and specificity to the story.}} There are \textcolor{red}{\underline{no apparent signs of}} \textcolor{red}{\underline{rebuttals}} or conflicts in the comments, and the tone remains supportive and positive. Considering the positive writing style, lack of conflicts in comments, and the specific details mentioned, the news appears more likely to be real. Therefore, the answer is: \textbf{B. Real}.\\
			\midrule
Final Output &
The final estimation of the veracity of the news is: \textbf{The news is real.}\\
			\bottomrule
	\end{tabular}}
        \caption{Sample ``3912024620676243'' in the Weibo dataset. LeRuD correctly predicts it as real news.}
	\label{tab:EX-3912024620676243}
\end{table*}

\begin{table*}[!ht]
	\centering
	\resizebox{\linewidth}{!}
	{\begin{tabular}{l|p{16cm}}
			\toprule
Prompt of Step 1 &
There is a piece of news: ``\textcolor{blue}{The Chinese Embassy in Mali has confirmed that three employees of China Railway Construction Corporation (CRCC) who went to Mali to negotiate a project with the Ministry of Transportation have died in the attack. They are Zhou Tianxiang, General Manager of China Railway Construction International Group, Wang Xuanshang, Deputy General Manager of China Railway Construction International Group, and Chang Xuehui, General Manager of China Railway Construction International Group West Africa Company.}'' There are comments for the news: ``\textcolor{blue}{Pray}'' ``\textcolor{blue}{There are risks overseas.}'' ``\textcolor{blue}{One of them met a year ago while eating in Cote, and their lives were so rushed. They were all people who struggled hard in Africa.}'' ``\textcolor{blue}{I'm too sad, I hope my compatriots can rest in peace soon.}'' ``\textcolor{blue}{The three Chinese citizens who died in the Mali hotel attack are employees of China Railway Construction Corporation!}'' ``\textcolor{blue}{The deceased rest in peace.}'' ``\textcolor{blue}{The risk of enterprises going global is increasing, and our responsibility to protect citizens overseas is becoming heavier. Our overseas military strike capabilities need to be greatly strengthened.}'' …… You need to do: (1) Based on the writing style and the commonsense knowledge, estimate the credibility of the news. (2) Based on the comments, analyze whether there are any rebuttals or conflicts, and then accordingly verify the authenticity of the news. Based on above results, please choose the answer from the following options: A. Fake, B. Real.\\
			\midrule
Output of Step 1 &
The news appears to be written in \textcolor{red}{\underline{a straightforward and factual manner}}, which adds to its credibility. \textcolor{red}{\underline{The details about the individuals involved, their positions, and the circumstances of the incident provide}} \textcolor{red}{\underline{a sense of specificity.}} The news itself \textcolor{red}{\underline{doesn't seem to conflict with common knowledge}}. Attacks and security risks can occur in various regions, including those with ongoing geopolitical tensions or internal conflicts. The comments express sympathy, sorrow, and concern for the individuals involved. There are \textcolor{red}{\underline{no apparent rebuttals or conflicts}} among the comments, which suggests a collective acknowledgment of the tragic event. Therefore, the answer is: \textbf{B. Real}.\\
\midrule
Prompt of Step 2 &
There is a piece of news: ``\textcolor{blue}{The Chinese Embassy in Mali has confirmed that three employees of China Railway Construction Corporation (CRCC) who went to Mali to negotiate a project with the Ministry of Transportation have died in the attack. They are Zhou Tianxiang, General Manager of China Railway Construction International Group, Wang Xuanshang, Deputy General Manager of China Railway Construction International Group, and Chang Xuehui, General Manager of China Railway Construction International Group West Africa Company.}'' There are comments for the news: ``\textcolor{blue}{Oh my, work is so dangerous.}'' ``\textcolor{blue}{I used to think these things were so far away from me. Since my brother went to Africa. Every time I see such news, my heart beats tightly, hoping for world peace!!!!}'' ``\textcolor{blue}{Three elites, three families.}'' ``\textcolor{blue}{Sadness, it's all corporate executives, keep going!}'' ``\textcolor{blue}{The reason why we cannot see darkness is because someone is doing their best to block it from where we cannot see. All warriors who fight against terrorist forces, regardless of nationality, faction, or stance, should be remembered by us!}'' ``\textcolor{blue}{I saw such heartbreaking news early in the morning!}'' ``\textcolor{blue}{Mourning for the deceased! Perhaps the child loses the father, the lover loses the husband, and the white haired person gives the black haired person away.}'' …… You need to do: (1) Based on the writing style and the commonsense knowledge, estimate the credibility of the news. (2) Based on the comments, analyze whether there are any rebuttals or conflicts, and then accordingly verify the authenticity of the news. Based on above results, please choose the answer from the following options: A. Fake, B. Real.\\
			\midrule
Output of Step 2 &
The news is presented in \textcolor{red}{\underline{a straightforward and factual manner}}. The \textcolor{red}{\underline{details about the victims and the}} \textcolor{red}{\underline{circumstances of the incident are specific}}, contributing to the credibility of the news. \textcolor{red}{\underline{The idea that work}} \textcolor{red}{\underline{in certain regions, especially those with geopolitical tensions or security concerns, can be dangerous ali-}} \textcolor{red}{\underline{gns with common knowledge.}} The news about individuals associated with a construction corporation facing risks during overseas negotiations is plausible. The comments express empathy, concern, and sadness for the victims and their families. There are \textcolor{red}{\underline{no apparent rebuttals or conflicts}} in the comments, suggesting a shared sentiment among the commenters regarding the tragic event. Therefore, the answer is: \textbf{B. Real}.\\
			\midrule
Final Output &
The final estimation of the veracity of the news is: \textbf{The news is real.}\\
			\bottomrule
	\end{tabular}}
        \caption{Sample ``3911539469801102'' in the Weibo dataset. LeRuD correctly predicts it as real news.}
	\label{tab:EX-3911539469801102}
\end{table*}

\begin{table*}[!ht]
	\centering
	\resizebox{\linewidth}{!}
	{\begin{tabular}{l|p{16cm}}
			\toprule
Prompt of Step 1 &
There is a piece of news: ``\textcolor{blue}{Man armed with knife on Swiss train starts fire - seven people hospitalised with knife wounds and burns, police say.}'' There are comments for the news: ``\textcolor{blue}{Mental Illness ?}'' ``\textcolor{blue}{They put on my wall and the nearby Antonia and neighboring Edvaldo one boma H and obvious that I wiretaps giant recorders}'' ``\textcolor{blue}{hope is not wat am thinking}'' ``\textcolor{blue}{To Germany he was working well with the H-bomb was the one who was present in the second war with his amateur radio.}'' ``\textcolor{blue}{please not another Machete!}'' ``\textcolor{blue}{ i don't even care anymore, terrorists is now a daily thing.}'' ``\textcolor{blue}{horrible attack.}'' …… You need to do: (1) Based on the writing style and the commonsense knowledge, estimate the credibility of the news. (2) Based on the comments, analyze whether there are any rebuttals or conflicts, and then accordingly verify the authenticity of the news. Based on above results, please choose the answer from the following options: A. Fake, B. Real.\\
			\midrule
Output of Step 1 &
Based on the provided information, it's challenging to directly assess the credibility of the news as there are no specific details about the source, and the content is limited to a brief summary. However, the news itself doesn't inherently raise red flags that would suggest it's fake. \textcolor{red}{\underline{News about violent incidents can occur, and it is not uncommon for such events to be reported.}} Analyzing the comments, it seems there is a mix of reactions ranging from speculation about mental illness, conspiracy theories about wiretaps, concerns about terrorism becoming a daily occurrence, and references to historical events. These comments do \textcolor{red}{\underline{not provide concrete information about the incident itself but rather express various opinions and thoug-}} \textcolor{red}{\underline{hts}} from different perspectives. Therefore, the answer is: \textbf{B. Real}.\\
\midrule
Prompt of Step 2 &
There is a piece of news: ``\textcolor{blue}{Man armed with knife on Swiss train starts fire - seven people hospitalised with knife wounds and burns, police say.}'' There are comments for the news: ``\textcolor{blue}{there are bad people in every ethical community}'' ``\textcolor{blue}{bloody hell, this is awful :( }'' ``\textcolor{blue}{hope is not wat am thinking}'' ``\textcolor{blue}{worst killers are mentally ill or have problems in their personal environment}'' ``\textcolor{blue}{yes i am afraid, there have been 28,000 Islamic terrorist attacks around the world since 2001}'' ``\textcolor{blue}{maybe take a look back in history the crusades were almost the same thing what Isis is doing now conquer land in the name of}'' ``\textcolor{blue}{Sorry!}'' …… You need to do: (1) Based on the writing style and the commonsense knowledge, estimate the credibility of the news. (2) Based on the comments, analyze whether there are any rebuttals or conflicts, and then accordingly verify the authenticity of the news. Based on above results, please choose the answer from the following options: A. Fake, B. Real.\\
			\midrule
Output of Step 2 &
Based on the provided information, the news about a man armed with a knife on a Swiss train starting a fire, resulting in seven people being hospitalized with knife wounds and burns, appears credible. The report is \textcolor{red}{\underline{specific and details the incident}}, which is \textcolor{red}{\underline{consistent with the style of legitimate news reporting}}. Analyzing the comments, there is a mix of reactions expressing concern, empathy, and discussions about the potential motives or background of the attacker. While there is a mention of Islamic terrorist attacks globally, it doesn't directly rebut or conflict with the reported incident on the Swiss train. The comments \textcolor{red}{\underline{mostly reflect individual opinions and reactions to the news rather than providing contradictory informa-}} \textcolor{red}{\underline{tion}}. Therefore, the answer is: \textbf{B. Real}.\\
			\midrule
Prompt of Step 3 &
There is a piece of news: ``\textcolor{blue}{Man armed with knife on Swiss train starts fire - seven people hospitalised with knife wounds and burns, police say.}'' There are comments for the news: ``\textcolor{blue}{learn to read dude. Maybe in the future, read before commenting...}'' ``\textcolor{blue}{Maybe another follower of the religion of peace that's invaded Europe.}'' ``\textcolor{blue}{hope is not wat am thinking}'' ``\textcolor{blue}{I agree that most of these people do pretend to follow this religion but they could also say they follow the}'' ``\textcolor{blue}{ Source please. I mean actual sources contributing unhealthy-ness to con/chem trails.}'' ``\textcolor{blue}{Pathetic}'' ``\textcolor{blue}{how very inconvenient}'' …… You need to do: (1) Based on the writing style and the commonsense knowledge, estimate the credibility of the news. (2) Based on the comments, analyze whether there are any rebuttals or conflicts, and then accordingly verify the authenticity of the news. Based on above results, please choose the answer from the following options: A. Fake, B. Real.\\
			\midrule
Output of Step 3 &
Based on the provided information, the news about a man armed with a knife on a Swiss train starting a fire, resulting in seven people being hospitalized with knife wounds and burns, appears credible. The report is \textcolor{red}{\underline{specific and details the incident}}, which is \textcolor{red}{\underline{consistent with the style of legitimate news reporting}}. Analyzing the comments, there is some indication of conflicting opinions or skepticism. Comments suggesting a connection to a specific religion and questioning the authenticity by asking for sources may imply some level of doubt or disagreement. However, it's important to note that \textcolor{red}{\underline{comments alone may not provide concrete evidence to dispute the credibility of the news}}. Therefore, the answer is: \textbf{B. Real}.\\
			\midrule
Final Output &
The final estimation of the veracity of the news is: \textbf{The news is real.}\\
			\bottomrule
	\end{tabular}}
        \caption{Sample ``764497123530375169'' in the Twitter dataset. LeRuD correctly predicts it as real news.}
	\label{tab:EX-764497123530375169}
\end{table*}

\begin{table*}[!ht]
	\centering
	\resizebox{\linewidth}{!}
	{\begin{tabular}{l|p{16cm}}
			\toprule
Prompt of Step 1 &
There is a piece of news: ``\textcolor{blue}{NASA astronaut captures a beautiful sunrise from aboard the International Space Station}'' There are comments for the news: ``\textcolor{blue}{heavenly sigh....}'' ``\textcolor{blue}{Beautiful}'' ``\textcolor{blue}{Wheres the UFO pictures?}'' ``\textcolor{blue}{this is cool as hell}'' ``\textcolor{blue}{Nature is beauty}'' ``\textcolor{blue}{Thanks to \#NASA  to capture such a scene !}'' ``\textcolor{blue}{woah it looks like the universal studios logo, the old one from 1990's?}'' …… You need to do: (1) Based on the writing style and the commonsense knowledge, estimate the credibility of the news. (2) Based on the comments, analyze whether there are any rebuttals or conflicts, and then accordingly verify the authenticity of the news. Based on above results, please choose the answer from the following options: A. Fake, B. Real.\\
			\midrule
Output of Step 1 &
The news itself seems plausible, as it describes a NASA astronaut capturing a sunrise from the International Space Station, which \textcolor{red}{\underline{aligns with the activities and capabilities of NASA}}. The language used is neutral and descriptive, \textcolor{red}{\underline{without any sensational or unrealistic claims}}. Additionally, the idea of astronauts capturing images from the ISS is a common and well-established practice. The \textcolor{red}{\underline{comments are generally}} \textcolor{red}{\underline{positive and express awe and appreciation}} for the captured sunrise. There doesn't seem to be any overt disagreement or conflicting opinions. However, one comment humorously asks about UFO pictures, but this seems more like \textcolor{red}{\underline{a playful remark rather than a serious rebuttal}}. \textcolor{red}{\underline{No significant rebuttals or conflicts}} found; comments are in line with the nature of the news. Therefore, the answer is: \textbf{B. Real}.\\
			\midrule
Final Output &
The final estimation of the veracity of the news is: \textbf{The news is real.}\\
			\bottomrule
	\end{tabular}}
        \caption{Sample ``693119705469587456'' in the Twitter dataset. LeRuD correctly predicts it as real news.}
	\label{tab:EX-693119705469587456}
\end{table*}

\begin{table*}[!ht]
	\centering
	\resizebox{\linewidth}{!}
	{\begin{tabular}{l|p{16cm}}
			\toprule
Prompt of Step 1 &
There is a piece of news: ``\textcolor{blue}{French police arrest woman who was with man carrying weapons and ammunition at Disneyland Paris hotel.}'' There are comments for the news: ``\textcolor{blue}{good. Paris doesn't play. I do not blame them.}'' ``\textcolor{blue}{honey grab some guns, ammo and the Quran and we're off to Disneyland}'' ``\textcolor{blue}{thank you.}'' ``\textcolor{blue}{it does matter- a lot!}'' ``\textcolor{blue}{The man was armed with the koran, which was the greater danger...}'' ``\textcolor{blue}{And was packing a fellow ostensibly of some ostensibly specific genealogy?}'' ``\textcolor{blue}{people never read anymore. They put he was carrying a Qur'an in the subheadline in the article. Two seconds people.}'' …… You need to do: (1) Based on the writing style and the commonsense knowledge, estimate the credibility of the news. (2) Based on the comments, analyze whether there are any rebuttals or conflicts, and then accordingly verify the authenticity of the news. Based on above results, please choose the answer from the following options: A. Fake, B. Real.\\
			\midrule
Output of Step 1 &
Based on the provided information, the news appears credible as it reports \textcolor{red}{\underline{a specific event}} involving the arrest of a woman with a man carrying weapons and ammunition at a Disneyland Paris hotel. The writing style is \textcolor{red}{\underline{straightforward and consistent with typical news reporting}} on law enforcement activities. Analyzing the comments, there are different perspectives expressed. Some comments express support for the arrest and appreciation for the vigilance in handling potential threats, while others inject humor or skepticism into the situation. \textcolor{red}{\underline{The variety of responses is not uncommon in online discussions}}, especially regarding sensitive topics like security incidents. There are \textcolor{red}{\underline{no clear rebuttals or conflicts}} within the comments that challenge the authenticity of the news. Instead, the comments reflect a range of opinions and reactions to the reported event. Therefore, the answer is: \textbf{B. Real}.\\
			\midrule
Final Output &
The final estimation of the veracity of the news is: \textbf{The news is real.}\\
			\bottomrule
	\end{tabular}}
        \caption{Sample ``692764799315398656'' in the Twitter dataset. LeRuD correctly predicts it as real news.}
	\label{tab:EX-692764799315398656}
\end{table*}

\begin{table*}[!ht]
	\centering
	\resizebox{\linewidth}{!}
	{\begin{tabular}{l|p{16cm}}
			\toprule
Prompt of Step 1 &
There is a piece of news: ``\textcolor{blue}{Wow. Anthony Johnson's knockout win is the 3rd-quickest finish in light heavyweight history (0:13).}'' There are comments for the news: ``\textcolor{blue}{WWE is still fake}'' ``\textcolor{blue}{that's a bad man !!}'' ``\textcolor{blue}{his name is rumble}'' ``\textcolor{blue}{Johnson is a true professional. Very polite and respectful; kudos to you}'' ``\textcolor{blue}{The real hands of stone.}'' ``\textcolor{blue}{worked for it. Well deserved}'' ``\textcolor{blue}{I can't believe there's been TWO others quicker than that!! TWO!!}'' …… You need to do: (1) Based on the writing style and the commonsense knowledge, estimate the credibility of the news. (2) Based on the comments, analyze whether there are any rebuttals or conflicts, and then accordingly verify the authenticity of the news. Based on above results, please choose the answer from the following options: A. Fake, B. Real.\\
			\midrule
Output of Step 1 &
Based on the information provided, it is challenging to definitively determine the credibility of the news. The news itself states \textcolor{red}{\underline{a specific fact}} about Anthony Johnson's knockout win being the 3rd-quickest finish in light heavyweight history (0:13). However, without additional context or sources, it's difficult to verify the accuracy of this claim. Analyzing the comments \textcolor{red}{\underline{doesn't seem to reveal any direct rebuttals or conflicts}} \textcolor{red}{\underline{regarding the authenticity}} of the news. \textcolor{red}{\underline{Most of the comments express admiration}} for Anthony Johnson's performance, acknowledge his nickname ``Rumble,'' and commend him for being a true professional. Some \textcolor{red}{\underline{express surprise at the speed of the knockout but without disputing the information}}. Therefore, the answer is: \textbf{B. Real}.\\
			\midrule
Final Output &
The final estimation of the veracity of the news is: \textbf{The news is real.}\\
			\bottomrule
	\end{tabular}}
        \caption{Sample ``767203096472719364'' in the Twitter dataset. LeRuD correctly predicts it as real news.}
	\label{tab:EX-767203096472719364}
\end{table*}

\begin{table*}[!ht]
	\centering
	\resizebox{\linewidth}{!}
	{\begin{tabular}{l|p{16cm}}
			\toprule
Prompt of Step 1 &
There is a piece of news: ``\textcolor{blue}{A 3-year-old daughter broke a cup and her parents drove away: the cup was broken and the child compensated for it.}'' There are comments for the news: ``\textcolor{blue}{Such parents are not worthy of having children!}'' ``\textcolor{blue}{Such parents are not worthy of having children!}'' ``\textcolor{blue}{Can this constitute the crime of abandonment?}'' ``\textcolor{blue}{My god!}'' ``\textcolor{blue}{I want to say that the little girl is very sad! Very sad}'' ``\textcolor{blue}{Oh my goodness... How much psychological shadow does this have to cause for children.}'' ``\textcolor{blue}{Sending such a child back to their parents will lead to more hardships in the future, so be careful not to abuse them!}'' …… You need to do: (1) Based on the writing style and the commonsense knowledge, estimate the credibility of the news. (2) Based on the comments, analyze whether there are any rebuttals or conflicts, and then accordingly verify the authenticity of the news. Based on above results, please choose the answer from the following options: A. Fake, B. Real.\\
			\midrule
Output of Step 1 &
The news appears to be \textcolor{red}{\underline{written in a straightforward manner}}, describing a situation where a 3-year-old broke a cup, and her parents drove away while she compensated for it. However, without additional details or context, it's challenging to determine the full credibility of the news. The emotional language used in the comments may suggest \textcolor{red}{\underline{a strong reaction from the audience}}, but it doesn't necessarily provide evidence regarding the authenticity of the incident. The comments express strong negative sentiments towards the parents, with phrases like ``not worthy of having children'' and concerns about potential psychological impact on the child. However, there are no comments presenting alternative viewpoints or challenging the authenticity of the news. The comments mostly \textcolor{red}{\underline{focus on expressing shock, sympathy for the child, and concern for her well-being}}. Therefore, the answer is: \textbf{B. Real}.\\
			\midrule
Final Output &
The final estimation of the veracity of the news is: \textbf{The news is real.}\\
			\bottomrule
	\end{tabular}}
        \caption{Sample ``3914951724449159'' in the Weibo dataset. LeRuD correctly predicts it as real news.}
	\label{tab:EX-3914951724449159}
\end{table*}

\end{document}